\documentclass[aps,prd,twocolumn,amsmath,showpacs,superscriptaddress,nofootinbib]{revtex4-1}
\usepackage{graphicx}
\usepackage{amsmath}
\usepackage{longtable}
\usepackage{float}
\usepackage{dcolumn}
\usepackage{bm}
\usepackage{appendix}
\usepackage{multirow}
\usepackage{color}
\usepackage{soul}
\usepackage[normalem]{ulem}

\usepackage{hyperref}

\hypersetup{colorlinks,linkcolor={blue},citecolor={blue},urlcolor={blue}}

\begin{document}

\title{Primordial gravitational waves from NANOGrav: a broken power-law approach}

\author{Micol Benetti}
\email{micol.benetti@unina.it}
\affiliation{Scuola Superiore Meridionale (SSM), Universit\`{a} di Napoli ``Federico II'', Largo San Marcellino 10, I-80138 Napoli, Italy}
\affiliation{Istituto Nazionale di Fisica Nucleare (INFN), Sezione di Napoli, Via Cinthia 9, I-80126 Napoli, Italy}

\author{Leila L. Graef}
\email{leilagraef@id.uff.br}
\affiliation{Instituto de F\'{i}sica, Universidade Federal Fluminense,
Avenida General Milton Tavares de Souza s/n, Gragoat\'{a}, 24210-346 Niter\'{o}i, Rio de Janeiro, Brazil}

\author{Sunny Vagnozzi}
\email{sunny.vagnozzi@ast.cam.ac.uk}
\affiliation{Kavli Institute for Cosmology (KICC) and Institute of Astronomy,\\University of Cambridge, Madingley Road, Cambridge CB3 0HA, United Kingdom}

\begin{abstract}
\noindent We revisit the possibility that the stochastic common-spectrum process recently detected by the NANOGrav pulsar timing array experiment could be due to primordial gravitational waves (GWs). A na\"{i}ve extrapolation down to interferometer scales of the blue GW spectrum required to explain NANOGrav consistently with Cosmic Microwave Background (CMB) observations would strongly violate upper limits on the stochastic GW background (SGWB) amplitude from LIGO/Virgo. In combination with the fact that there are over 19 decades in frequency between CMB and interferometer scales, this motivates us to move beyond the commonly adopted approximation of a pure power-law GW spectrum. We consider a broken power-law parametrization for the SGWB spectrum, which turns from blue to red above the break frequency: while phenomenological, this choice maps to various well-motivated early-Universe models, including scenarios featuring non-instantaneous reheating or a non-standard background expansion following reheating. After a detailed discussion of the contribution of the resulting SGWB to the early-Universe radiation energy density, we constrain the broken power-law model against a wide variety of multi-frequency cosmological and GW observations. We find that this phenomenological model is able to explain the NANOGrav signal while remaining in agreement with upper limits on the tensor-to-scalar ratio on CMB scales, Big Bang Nucleosynthesis constraints on the early-Universe radiation energy density, and upper limits on the SGWB amplitude on interferometer scales. We briefly discuss the very bright prospects for testing this model with next-generation probes across the GW frequency landscape, which motivate further exploring connections to specific well-motivated early-Universe models.

\end{abstract}

\date{\today}

\maketitle

\section{Introduction}
\label{sec:intro}

The first detection of gravitational waves (GWs) from the coalescence of two black holes (BHs) in 2015~\cite{LIGOScientific:2016aoc} has opened an entirely new window onto the Universe, the behavior of its most extreme objects, and the laws of physics at energy scales and regimes which are completely unaccessible on Earth~\cite{Sathyaprakash:2009xs,Cai:2017cbj,Bian:2021ini}. While so far only resolved GW events have been detected, the superposition of numerous incoherent GW sources would instead generate a stochastic GW background (SGWB). The existence of a SGWB is a firm prediction of several well-motivated cosmological and astrophysical scenarios, operating both in the early and late Universe (see e.g. Refs.~\cite{Maggiore:1999vm,Caprini:2018mtu,Giovannini:2019oii} for reviews). Merging supermassive BH (SMBH) binaries are perhaps the best motivated example on the astrophysical side. On the cosmological side, SGWB sources include for instance phase transitions and relics thereof (including topological defects such as cosmic strings). There is no doubt that a direct observation of the cosmological or astrophysical SGWB would be a tremendous achievement, whose implications for astrophysics, cosmology, and high-energy physics would be momentous. 

The predicted astrophysical and cosmological SGWB spans an extremely wide frequency range. Significant theoretical and experimental effort has gone into the development of a diverse range of probes (currently running, upcoming, or proposed) which will be able to search for GWs (both resolved GW events or the SGWB) in various frequency bands, from low-frequency ($f \sim 10^{-20}\,{\rm Hz}$) GWs to GWs in the ${\rm kHz}$ band. From the lowest to the highest frequencies, these probes and/or observables include (but are not limited to) Cosmic Microwave Background (CMB) B-modes~\cite{Kamionkowski:2015yta}, spectral distortions~\cite{Ota:2014hha,Kite:2020uix}, pulsar timing arrays (PTAs)~\cite{1978SvA....22...36S,Detweiler:1979wn,1990ApJ...361..300F}, binary resonance~\cite{Blas:2021mpc,Blas:2021mqw}, and finally direct detection with atomic~\cite{Graham:2017pmn,AEDGE:2019nxb,Ellis:2020lxl}, Earth-~\cite{Abramovici:1992ah}, and space-based interferometers~\cite{Barausse:2020rsu}. Of particular interest to this work are PTAs, which are sensitive to GWs with frequencies in the ${\rm nHz}$ range, and exploit the fact that millisecond pulsar behave as extremely stable clocks. GWs induce spatially correlated fluctuations in the arrival times of radio pulses from millisecond pulsars~\cite{Yokoyama:2021hsa}, which PTAs then search for. The $\gtrsim{\cal O}({\rm nHz})$ region is of particular interest from the astrophysical and cosmological points of view: the astrophysical SGWB from merging SMBH binaries (SMBHBs) is expected to peak within this band~\cite{Sesana:2004sp}, and the same is true for the cosmological SGWB arising within some of the simplest models of cosmic strings~\cite{Blanco-Pillado:2017rnf}.

A particularly well-motivated cosmological SGWB source is cosmic inflation~\cite{Kazanas:1980tx,Starobinsky:1980te,Sato:1981ds,Guth:1980zm,Mukhanov:1981xt,Linde:1981mu,Albrecht:1982wi}, the leading paradigm for the solution of the flatness, horizon, and monopole problems, as well as the generation of primordial density perturbations. Inflationary vacuum fluctuations become classical on large scales, and induce both scalar and tensor perturbations. Upon horizon re-entry, the latter give rise to the inflationary SGWB, imprinting a distinctive signature in the CMB B-mode polarization, which is therefore among the cleanest probes of the inflationary SGWB~\cite{Kamionkowski:2015yta}. The amplitude and scale dependence of the inflationary SGWB is typically parametrized via the tensor-to-scalar ratio $r$ (characterizing the amplitude of tensor fluctuations relative to scalar ones) and the tensor spectral index $n_T$ respectively. The simplest models of inflation, driven by a single dynamical slowly-rolling (scalar) field, predict a spectrum of scalar fluctuations which is nearly scale-invariant (albeit slightly red, with more power on large rather than small scales) and highly Gaussian. These predictions are in excellent agreement with observations, which strongly constrain non-Gaussianity and deviations from scale-invariance, lending very strong support to the inflationary paradigm~\cite{Planck:2018jri}.

Single-field slow-roll models also predict a power spectrum of tensor fluctuations which is slightly red. To leading order in slow-roll parameters, the tensor-to-scalar ratio and tensor spectral index within these models satisfy the so-called inflationary consistency relation~\cite{Liddle:1993fq}:
\begin{eqnarray}
r = -8n_T\,,
\label{eq:consistency}
\end{eqnarray}
which thus requires $n_T \leq 0$ (hence a red spectrum), since $r \geq 0$. Within these models, and given current constraints on $r$~\cite{BICEP2:2018kqh}, the amplitude of the inflationary SGWB on PTA and interferometer scales is far too small to be detectable by these probes, which would instead require a strong blue tilt or in any case a strong enhancement on small scales. However, as we will discuss in more detail in Sec.~\ref{sec:spectra}, several well-motivated inflationary (and non-inflationary) models beyond the simplest ones naturally predict a blue tilt for the tensor power spectrum (see e.g. Refs.~\cite{Kobayashi:2010cm,Myrzakulov:2015qaa,Fujita:2018ehq,Kawai:2021edk,Oikonomou:2021kql,Odintsov:2021kup,Calcagni:2004as,Calcagni:2013lya,Endlich:2012pz,Cannone:2014uqa,Graef:2015ova,Ricciardone:2016lym,Graef:2017cfy,Baldi:2005gk,Maleknejad:2011sq,Adshead:2012kp,Maleknejad:2016qjz,Dimastrogiovanni:2016fuu,Adshead:2016omu,Obata:2016oym,Iacconi:2020yxn,Cook:2011hg,Pajer:2013fsa,Mukohyama:2014gba,Gruzinov:2004ty,Ashoorioon:2014nta,Giare:2020plo,Biagetti:2013kwa,Cai:2016ldn,Cai:2020ovp,Brandenberger:1988aj,Brandenberger:2006vv,Brandenberger:2006xi,Stewart:2007fu,Brandenberger:2014faa,Khoury:2001wf,Hipolito-Ricaldi:2016kqq,Brandenberger:2009jq}). Before moving on, let us mention that there is certainly ample theoretical motivation for going beyond the observationally highly successful single-field slow-roll paradigm. For instance, recent work within the ``swampland'' program has pointed out difficulties in embedding the simplest inflationary models within quantum gravity-consistent UV completions~\cite{Obied:2018sgi,Agrawal:2018own,Garg:2018reu,Ooguri:2018wrx,Bedroya:2019snp}, whereas such difficulties may be evaded when moving past the single-field slow-roll paradigm~\cite{Achucarro:2018vey,Kehagias:2018uem,Matsui:2018bsy,Kinney:2018nny,Brahma:2018hrd,Das:2018hqy,Motaharfar:2018zyb,Ashoorioon:2018sqb,Kinney:2018kew,Geng:2019phi,Odintsov:2020sqy,Odintsov:2020zkl,Trivedi:2020wxf,Oikonomou:2020oex,Trivedi:2021nss,Kawai:2021bye}. Therefore, the possibility that non-minimal inflationary models or alternatives to inflation leading to a blue tensor spectrum may be probed at PTA or interferometer scales is one which is highly worthy of consideration from both the theoretical and observational points of view. This will be the starting point for our work which, as anticipated earlier, shall be concerned with the SGWB in the $\gtrsim {\cal O}({\rm nHz})$ frequency range, to which PTAs are sensitive.

The North American Nanohertz Observatory for Gravitational Waves (NANOGrav) is a PTA collaboration which has been collecting pulsar timing data since 2004~\cite{2019BAAS...51g.195R}. NANOGrav recently released their 12.5-year dataset~\cite{NANOGrav:2020gpb}, which contains time-of-arrival measurements for 47 millisecond pulsars observed at the Arecibo Observatory and the Green Bank Telescope between 2004 and 2017. An analysis searching for an isotropic SGWB by the NANOGrav collaboration in their 12.5-year dataset yielded strong evidence for a stochastic common-spectrum process against independent red-noise processes~\cite{NANOGrav:2020bcs}, with Bayes factors in found to lie within the range $2.7 \lesssim \log_{10}B \lesssim 4.5$ depending on the solar system ephemeris (SSE) modeling scheme adopted. This result, if genuine and confirmed, raises the tantalizing possibility that NANOGrav may have achieved the first ever SGWB detection. The NANOGrav signal is consistent with a SGWB with characteristic strain amplitude $A_{\rm CP} \sim 10^{-15}$~\footnote{The subscript ``CP'' stands for ``common(-spectrum) process''.} at a reference frequency $f_{\rm yr}=1\,{\rm yr}^{-1} \approx 3 \times 10^{-8}\,{\rm Hz}$.

It is worth pointing out that the NANOGrav detection shows no evidence for quadrupolar spatial correlations~\cite{NANOGrav:2020bcs}, as described by the Hellings-Downs (HD) curve~\cite{Hellings:1983fr}.~\footnote{The Bayes factor against a spatially uncorrelated common-spectrum process lies within the range $0.37 \lesssim \log_{10}B \lesssim 0.64$, again depending on the SSE model adopted~\cite{NANOGrav:2020bcs}.} The lack of evidence for ``tell-tale'' quadrupolar HD correlations constitutes grounds for caution with regards to claims of the NANOGrav signal constituting a genuine SGWB detection, and the NANOGrav collaboration themselves advocate for a prudent approach on the matter~\cite{NANOGrav:2020bcs}. Intriguingly, the Parkes Pulsar Timing Array (PPTA) collaboration very recently also reported evidence for a common-spectrum process with amplitude and frequency range consistent with the NANOGrav signal in their DR2 dataset~\cite{Goncharov:2021oub}. However, this tentative signal once again lacks convincing evidence for quadrupolar HD correlations. Even more recently, similar hints (though once more including the lack of quadrupolar HD correlations) were reported in the European Pulsar Timing Array (EPTA) 24-year dataset~\cite{Chen:2021rqp}, strengthening the hints for the possible first detection of a SGWB in the ${\rm nHz}$ range.~\footnote{The possibility that NANOGrav may have observed GW polarization modes other than the standard tensor transverse (TT) ones present in General Relativity (GR) was first discussed in detail in Ref.~\cite{Chen:2021wdo}. The authors claimed strong evidence for scalar transverse (ST) polarization modes, and no evidence for scalar longitudinal (SL) or vector longitudinal (VL) polarization modes. A similar study was performed in Ref.~\cite{Chen:2021ncc} on data from the International Pulsar Timing Array (IPTA), a collaboration comprising the EPTA in addition to NANOGrav and PPTA~\cite{Hobbs:2009yy}, finding this time weak evidence for ST polarization modes, and placing upper limits on the amplitude of TT polarization modes consistent with the NANOGrav signal. Yet another related study was performed in Ref.~\cite{Wu:2021kmd} on PPTA data, finding no evidence for either of the TT, ST, SL, and VL polarization modes, but only placing upper limits on the amplitude thereof, posing questions as to the origin of the tentative detection of ST polarization modes in earlier works. Finally, the NANOGrav collaboration themselves searched for non-GR polarization modes in Ref.~\cite{NANOGrav:2021ini}, again finding a preference for ST modes over TT ones, while showing that including modeling of SSE systematics and/or removing pulsar J0030+0451 from the analysis reduces the significance of the ST modes detection.}

Keeping in mind the important HD-related caveats discussed above, it is nevertheless worthwhile to contemplate possible scenarios explaining the NANOGrav signal, assuming it constitutes a genuine SGWB detection. As mentioned earlier, the astrophysical SGWB from merging SMBHBs is expected to peak within the range probed by NANOGrav, and remains a possible valid interpretation of the signal~\cite{Middleton:2020asl}, although the required local number density of SMBHBs is a factor of five larger than predicted by most astrophysical models~\cite{Casey-Clyde:2021xro}. Quite predictably, the NANOGrav detection has stimulated significant activity focused on identifying possible fundamental physics scenarios responsible for the signal, mostly operating in the primordial Universe. Some of the earliest interpretations of the NANOGrav signal involved cosmic strings~\cite{Ellis:2020ena,Blasi:2020mfx,Buchmuller:2020lbh,Samanta:2020cdk} or second-order effects associated to the formation of primordial BHs from the collapse of large curvature perturbations~\cite{Vaskonen:2020lbd,DeLuca:2020agl}, but various other theoretical scenarios have been considered since then (see e.g. Refs.~\cite{Addazi:2020zcj,Ratzinger:2020koh,Nakai:2020oit,Neronov:2020qrl,Li:2020cjj,Kohri:2020qqd,Sugiyama:2020roc,Namba:2020kij,Zhou:2020kkf,Liu:2020mru,Paul:2020wbz,Domenech:2020ers,Bhattacharya:2020lhc,Abe:2020sqb,Kitajima:2020rpm,Inomata:2020xad,Tahara:2020fmn,Chigusa:2020rks,Pandey:2020gjy,Bigazzi:2020avc,Ramberg:2020oct,Cai:2020qpu,Barman:2020jrf,Chiang:2020aui,Atal:2020yic,Datta:2020bht,Gao:2021dfi,Li:2021qer,Gorghetto:2021fsn,Kawasaki:2021ycf,Blanco-Pillado:2021ygr,Brandenburg:2021tmp,Hindmarsh:2021mnl,Lazarides:2021uxv,Zhou:2021cfu,Sakharov:2021dim,NANOGrav:2021flc,Borah:2021ocu,Yi:2021lxc,Wu:2021zta,Haque:2021dha,Liu:2021svg,Buchmuller:2021mbb,Masoud:2021prr,Li:2021htg,Spanos:2021hpk,Khodadi:2021ees,Izquierdo-Villalba:2021prf,Gao:2021lno,Lin:2021vwc} for examples, as well as Ref.~\cite{Bian:2020urb} for a comprehensive characterization of various possible sources for the NANOGrav signal).

Another intriguing possibility, which was first explored by one of us in Ref.~\cite{Vagnozzi:2020gtf}, and then further in Ref.~\cite{Kuroyanagi:2020sfw}, posits that the NANOGrav may be due to an inflationary SGWB. In particular, under the commonly adopted assumption of a pure power-law parametrization for the underlying primordial tensor power spectrum, Ref.~\cite{Vagnozzi:2020gtf} finds that explaining the NANOGrav signal requires: \textit{a)} a very blue ($n_T>0$) primordial tensor power spectrum, and hence a violation of the consistency relation in order to remain consistent with upper limits on the tensor-to-scalar ratio (and therefore the SGWB amplitude) on CMB scales; \textit{b)} a very low reheating temperature, in order to not violate Big Bang Nucleosynthesis (BBN) constraints on the radiation energy density in the early Universe. Related findings were also reported in Ref.~\cite{Kuroyanagi:2020sfw}, where late-time entropy production between the end of inflation and the BBN epoch was also taken into consideration. The possibility that the NANOGrav signal may be due to inflationary GWs remains therefore viable, but would require an inflationary model beyond the simplest single-field slow-roll ones, and a non-standard reheating and/or post-reheating scenario.

In this work, we shall keep pursuing the intriguing possibility that the NANOGrav collaboration may indeed have detected a primordial SGWB, remaining agnostic as to whether the latter is related or not to inflation. Our analysis goes beyond these earlier works (and in particular Ref.~\cite{Vagnozzi:2020gtf}) in at least three important respects:
\begin{itemize}
\item we go beyond widely adopted, but overly simplistic, approximation of a pure power-law for the primordial SGWB spectrum;
\item in addition to BBN bounds, we also take into account SGWB constraints at higher frequencies from LIGO/Virgo~\cite{LIGOScientific:2016jlg,LIGOScientific:2019vic}, which the model studied in Ref.~\cite{Vagnozzi:2020gtf} would na\"{i}vely violate;
\item we perform a fully-fledged likelihood analysis including (besides NANOGrav) other standard cosmological datasets to more properly constrain the primordial SGWB interpretation of the NANOGrav signal, and paying particular attention to the Hubble constant $H_0$, whose inferred value can be particularly sensitive to the existence of a blue GW spectrum, given the contribution of the latter to the effective number of relativistic species $N_{\rm eff}$, which correlates with $H_0$.
\end{itemize}
The first of the above three points is particularly important. While the approximation of SGWB power spectrum being described by a pure power-law across various decades in frequency is widespread, it might not be justified when considering probes spanning a wide frequency range and hence providing a large lever arm~\cite{Kinney:2021nje}: for instance, there are 19 decades in frequency between the scales probed by CMB experiments such as \textit{Planck} and interferometers such as LIGO/Virgo. Here, we shall go beyond the power-law form for the primordial tensor power spectrum, adopting a phenomenological broken power-law ansatz, recently proposed in Ref.~\cite{Kuroyanagi:2014nba}. While phenomenological, this parametrization (which we shall refer to as ``broken power-law spectrum'') covers a broad range of early-Universe scenarios, ranging from non-instantaneous reheating to a non-standard post-reheating background evolution.

The rest of this paper is then organized as follows. In Sec.~\ref{sec:spectra} we introduce the standard power-law SGWB spectrum and the associated energy density and number of relativistic species $N_{\rm eff}$, before presenting the broken power-law SGWB power spectrum and discussing the associated observational quantities such as $N_{\rm eff}$. In Sec.~\ref{sec:sources} we introduce the GW observations against which we constrain the model, placing constraints on the SGWB energy density within various frequency ranges. In Sec.~\ref{sec:analysis} we outline the cosmological data and analysis methodology adopted. Our results, and in particular constraints on the broken power-law SGWB spectrum and the possibility of the NANOGrav signal being due to primordial GWs, are presented in Sec.~\ref{sec:results}. Finally, we draw concluding remarks in Sec.~\ref{sec:conclusions}.

\section{Standard and non-standard primordial GW spectra}
\label{sec:spectra}

\subsection{Standard power-law spectrum}
\label{subsec:power-law}

The primordial spectrum of tensor perturbations ${\cal P}_T^{\rm prim}(k)$ is a key quantity to connect theoretical early-Universe predictions (including those from inflation) to cosmological observations. A widespread choice in the literature is that of assuming that the primordial tensor spectrum scales as a pure power-law across various decades in comoving wavenumber $k$:
\begin{eqnarray}
{\cal P}_T^{\rm prim,pl}(k) = A_T(k_{\star}) \left ( \frac{k}{k_{\star}} \right ) ^{n_T}\,,
\label{eq:ptprimpowerlaw}
\end{eqnarray}
where $A_T$ is the amplitude of the primordial tensor power spectrum (itself related to the tensor-to-scalar ratio $r \equiv A_T/A_s$, where $A_s$ is instead the amplitude of the primordial power spectrum of scalar perturbations) at the pivot scale $k_{\star}$, $n_T$ is the tensor tilt assumed to be a constant throughout the spectrum, and the superscript ``pl'' stands for ``power-law''. It can be convenient to express the above spectrum in frequency space, with frequency $f$ related to $k$ by $k=2\pi f$.

Within single-field slow-roll inflationary models, the consistency relation implies $r=-8n_T$, and therefore a negative tensor spectral index and correspondingly a red GW spectrum. However, as alluded to earlier, this is not necessarily the case in a number of other (less minimal but not for this reason less motivated) early-Universe scenarios (particularly those involving violations of the null-energy condition), including but not limited to: inflationary models based on modifications to gravity~\cite{Kobayashi:2010cm,Myrzakulov:2015qaa,Fujita:2018ehq,Kawai:2021edk,Oikonomou:2021kql,Odintsov:2021kup}, non-commutative space-times~\cite{Calcagni:2004as,Calcagni:2013lya}, spatial or temporal diffeomorphism invariance breaking~\cite{Endlich:2012pz,Cannone:2014uqa,Graef:2015ova,Ricciardone:2016lym,Graef:2017cfy}, and more generally violations of the null-energy condition~\cite{Baldi:2005gk}, or models involving couplings to gauge fields and spin-2 fields~\cite{Maleknejad:2011sq,Adshead:2012kp,Maleknejad:2016qjz,Dimastrogiovanni:2016fuu,Adshead:2016omu,Obata:2016oym,Iacconi:2020yxn}, particle production during inflation~\cite{Cook:2011hg,Pajer:2013fsa,Mukohyama:2014gba}, elastic media~\cite{Gruzinov:2004ty}, a non-Bunch-Davies initial state~\cite{Ashoorioon:2014nta}, higher order effective gravitational action corrections~\cite{Giare:2020plo}, second-order effects~\cite{Biagetti:2013kwa} (possibly associated to the formation of primordial BHs), sound speed resonance or a decrease in the GW sound speed during inflation~\cite{Cai:2016ldn,Cai:2020ovp}, as well as alternatives to inflation such as string gas cosmology~\cite{Brandenberger:1988aj,Brandenberger:2006vv,Brandenberger:2006xi,Stewart:2007fu,Brandenberger:2014faa}, ekpyrotic scenarios~\cite{Khoury:2001wf,Hipolito-Ricaldi:2016kqq}, and matter bounces~\cite{Brandenberger:2009jq}.~\footnote{See e.g. Ref.~\cite{Wang:2014kqa} for a comprehensive discussion of inflationary models and alternatives leading to a blue spectrum, motivated by BICEP2's claimed detection in 2014~\cite{BICEP2:2014owc,Gerbino:2014eqa}.}

The tensor power spectrum ${\cal P}_T(\eta,k)$ at a given conformal time $\eta$ is related to its primordial counterpart ${\cal P}_T^{\rm prim}(k)$ as follows:
\begin{eqnarray}
{\cal P}_T(\eta,k) = {\cal T}_T^2(\eta,k){\cal P}_T^{\rm prim}(k)\,,
\label{eq:ptpowerlaw}
\end{eqnarray}
where ${\cal T}_T(\eta,k)$ is the transfer function, which accounts for the evolution of tensor perturbations across the various epochs of the Universe's expansion history up to $\eta$. Assuming that reheating at the end of inflation is followed by the standard epochs of radiation domination, matter domination, and dark energy domination, the transfer function admits a simple analytical approximation (see Refs.~\cite{Turner:1993vb,Chongchitnan:2006pe,Zhao:2006mm,Nakayama:2008wy,Nakayama:2009ce,Kuroyanagi:2011fy,Zhao:2013bba}), which we too shall adopt.~\footnote{See Ref.~\cite{Kite:2021yoe} for a recent re-appraisal of some of these widely adopted approximations.}

Another key quantity in connecting theory to observations is the GW energy density today $\rho_{\rm GW}$, given by:
\begin{eqnarray}
\rho_{\rm GW}= \int^{k_{\rm{UV}}}_{k_{\rm{IR}}} d\ln k\,\frac{{\cal P}_T(k)}{32 \pi G a^{2}} \left [ {\cal T}'(k,\eta_0) \right ]^{2}\,,
\label{eq:rhogw}
\end{eqnarray}
where $\eta_0$ is the current conformal time, and $'$ denotes a conformal time derivative. The upper and lower integration limits $k_{\rm UV}$ and $k_{\rm IR}$ correspond to physical ultraviolet (UV) and infrared (IR) cutoff scales. For the purpose of comparison to observations, it is also convenient to define the GW dimensionless density parameter $\Omega_{\rm GW}$, given by (see e.g. Refs.~\cite{Watanabe:2006qe,Giare:2020plo,Dimastrogiovanni:2021mfs}):
\begin{eqnarray}
\Omega_{\rm GW} \equiv \frac{1}{\rho_c}\frac{d\rho_{\rm GW}}{d\ln k} = \frac{1}{12} \left ( \frac{k}{a_0H_0} \right )^2{\cal T}_T^2(k){\cal P}_T^{\rm prim}(k)\,,
\label{eq:omegagw}
\end{eqnarray}
where $H_0$ and $a_0$ are the current values of the Hubble expansion rate and scale factor respectively, and $\rho_c = 3H_0^2/8\pi G$ is the critical density of the Universe today. In particular, we shall denote by $\Omega_{\rm GW}^{\rm pl}$ the GW density parameter (as a function of frequency) associated to the pure power-law primordial tensor spectrum ${\cal P}_T^{\rm prim,pl}(k)$ given in Eq.~(\ref{eq:ptprimpowerlaw}), \textit{i.e.} obtained combining Eqs.~(\ref{eq:ptprimpowerlaw},\ref{eq:omegagw}).

Mathematically speaking, the IR and UV cutoffs are introduced since the integral in Eq.~(\ref{eq:rhogw}) would diverge in the IR for $n_T \leq -4$ and in the UV for $n_T \geq -2$. Physically speaking, for the IR cutoff, the only modes which contribute to the radiation energy density at any given time are subhorizon modes, as those are the ones which oscillate and propagate as massless modes, hence contributing to the local energy density (although this statement is to some extent ambiguous, as the total energy density is a quantity which can only be measured averaging over several wavelengths). This implies that $k_{\rm IR}$ is a time-dependent quantity, as the horizon itself is time-dependent. In this work we shall be concerned with a blue GW spectrum, for which it is the UV modes which dominate the energy density integral. Therefore, we can safely take the limit $k_{\rm IR} \to 0\,{\rm Hz}$, or more precisely the limit $k_{\rm IR}/k_{\rm UV} \to 0$, with virtually no impact on our results (see Ref.~\cite{Meerburg:2015zua} for related discussions).

The choice of UV cutoff is, instead, more arbitrary, although our being interested in a blue GW spectrum implies that this choice will nonetheless have an important impact on our results. If inflation is responsible for the production of primordial tensor modes, one expects an UV cutoff given the size of the horizon at the end of inflation. For GUT-scale inflation and instant reheating, this would correspond approximately to $k_{\rm UV} \sim 10^{23}\,{\rm Mpc}^{-1}$ and hence $f_{\rm UV} \sim 10^9\,{\rm Hz}$~\cite{Cabass:2015jwe}. However, a more conservative assumption could be that of not committing to any specific early-Universe model for generating GWs, but simply to require that the power-law spectrum extends over $\sim$60 \textit{e}-folds, \textit{i.e.} the maximum amount of hot Big Bang expansion. For a pivot scale of $k_{\star}=10^{-2}\,{\rm Mpc}^{-1}$, this implies $k_{\rm UV}/k_{\star} \sim 10^{24}$, leading to $k_{\rm UV} \sim 10^{22}\,{\rm Mpc}^{-1}$ and therefore $f_{\rm UV} \sim 10^8\,{\rm Hz}$, a result which is very close to that obtained from the previous inflation-based argument~\cite{Meerburg:2015zua}.

Less conservative arguments could be used to justify higher values of $f_{\rm UV}$. For instance, one would in any case expect that the largest possible wavenumber for GWs produced in the early Universe is set by the Planck scale, for which $k_{\rm UV} \sim 10^{57}\,{\rm Mpc}^{-1}$ and accordingly $f_{\rm UV} \sim 10^{43}\,{\rm Hz}$. This choice was adopted, for instance, in Ref.~\cite{Stewart:2007fu}. A slightly less aggressive choice could be to replace the Planck scale by the GUT scale, from which one gets $k_{\rm UV} \sim 10^{54}\,{\rm Mpc}^{-1}$ and $f_{\rm UV} \sim 10^{40}\,{\rm Hz}$. Of course, a sharp cut in $k$-space is a simplification and one would in fact generically expect a smooth transition between modes which contribute to the energy density and modes which don't. However, we do not expect this choice to have a significant impact on our results. We will discuss our specific choice for $f_{\rm UV}$ later in Sec.~\ref{subsec:brokenpowerlaw}, when discussing the choice of broken power-law spectrum we adopt, going beyond the pure power-law in Eq.~(\ref{eq:ptprimpowerlaw}).

For a blue GW spectrum ($n_T>0$), the integral in Eq.~(\ref{eq:rhogw}) can be solved by adopting the aforementioned analytical approximations for the transfer function ${\cal T}(\eta,k)$. Up to corrections of order $k_{\rm UV}/k_{\rm IR}$, which are extremely tiny, this gives (see e.g. Ref.~\cite{Meerburg:2015zua} for the full calculation):
\begin{eqnarray}
\rho_{\rm GW} &=& \frac{A_sr}{32\pi G} \left ( \frac{k_{\rm UV}}{k_{\star}} \right ) ^{n_T}\frac{1}{2n_T(a\eta)^2} \nonumber \\
&=& \frac{A_sr}{24n_T} \left ( \frac{k_{\rm UV}}{k_{\star}} \right ) ^{n_T}\rho_{\rm tot}\,,
\label{eq:rhogwpowerlaw}
\end{eqnarray}
where the second line is only valid deep during the radiation domination era, and follows from the fact that during this epoch $1/(a\eta)^2=H^2=8\pi G\rho_{\rm tot}/3$. Here, $\rho_{\rm tot}$ is the total energy density of the Universe, which during radiation domination is given by the sum of the photon ($\gamma$), neutrino ($\nu$), and GW energy densities:
\begin{eqnarray}
\rho_{\rm tot} &=&  \rho_\gamma + \rho_{\nu} + \rho_{\rm GW} \equiv \rho_{\gamma} \left ( 1 + \frac{7}{8}\left(\frac{4}{11} \right ) ^{4/3}N_{\rm eff} \right ) \nonumber\\
&=& \rho_{\gamma} \left ( 1 + \frac{7}{8}\left(\frac{4}{11}\right)^{4/3} 3.046 \right ) + \rho_{\rm GW}\,,
\label{eq:rhotot}
\end{eqnarray}
where the first line defines the effective number of relativistic degrees of freedom $N_{\rm eff}$ (including contributions from photons, neutrinos, and GWs), and in the second line we are assuming that the three Standard Model neutrino families provide the standard contribution $N_{\rm eff}^{\nu}=3.046$~\cite{Mangano:2005cc}.~\footnote{This standard value has recently been re-evaluated by several groups to include a more precise treatment of flavour oscillations and finite-temperature effects with the final results converging towards $N_{\rm eff}^{\nu} \simeq 3.044$ (see e.g. Refs.~\cite{deSalas:2016ztq,Gariazzo:2019gyi,Bennett:2019ewm,EscuderoAbenza:2020cmq,Akita:2020szl,Froustey:2020mcq,Bennett:2020zkv}). While we adopt the older standard value $N_{\rm eff}=3.046$, we note that adopting the latest value would have virtually no impact on our results, given the precision of current cosmological data, while even the precision of near-future cosmological data will still be more than one order of magnitude worse than $\Delta N_{\rm eff} \sim 0.002$~\cite{CMB-S4:2016ple,SimonsObservatory:2018koc,SimonsObservatory:2019qwx}.} Therefore, the total effective number of relativistic degrees of freedom is given by $N_{\rm eff}=N_{\rm eff}^{\nu}+N_{\rm eff}^{\rm GW}=3.046+N_{\rm eff}^{\rm GW}$, which defines the GW contribution to $N_{\rm eff}$, given by $N_{\rm eff}^{\rm GW}$.

If the GW energy density is a subdominant component of the total radiation energy density, i.e. $\rho_{\rm GW}/\rho_{\rm tot} \ll 1$, we can substitute Eq.~(\ref{eq:rhotot}) into Eq.~(\ref{eq:rhogwpowerlaw}), and solve for $N_{\rm eff}$ to obtain an expression which depends only on $A_s$, $r$, $n_T$, and $k_{\rm UV}$ (see Ref.~\cite{Meerburg:2015zua}). Taylor expanding to first order in $\rho_{\rm GW}/\rho_{\rm tot}$ leads to the following expression for $N_{\rm eff}$:
\begin{eqnarray}
N_{\rm eff} \approx 3.046 + \left [ 3.046 + \frac{8}{7} \left ( \frac{11}{4}\right)^{4/3} \right ] \frac{A_sr}{24n_T} \left ( \frac{k_{\rm UV}}{k_{\star}} \right )^{n_T}\,,\nonumber \\
\label{eq:neffpowerlaw}
\end{eqnarray}
an expression which we stress again is valid only if the GW energy density is small compared to the total radiation energy density, which fortunately is the case given observational constraints. From Eq.~(\ref{eq:neffpowerlaw}) we can directly read off $N_{\rm eff}^{\rm GW}$, given by:
\begin{eqnarray}
N_{\rm eff}^{\rm GW} \approx \left [ 3.046 + \frac{8}{7} \left ( \frac{11}{4}\right)^{4/3} \right ] \frac{A_sr}{24n_T} \left ( \frac{k_{\rm UV}}{k_{\star}} \right )^{n_T}\,.
\label{eq:neffgwpowerlaw}
\end{eqnarray}
which we recall is only valid for a blue GW spectrum.

In the literature, the assumption of the primordial tensor power spectrum scaling as a pure power-law up to the cutoff frequency $k_{\rm UV}$ as in Eq.~(\ref{eq:ptprimpowerlaw}) is a widespread one: see e.g. Refs.~\cite{Meerburg:2015zua,Escudero:2015wba,Cabass:2015jwe,Huang:2015cke,Bartolo:2016ami,Wang:2016tbj,Huang:2017gmp,Graef:2018fzu,Li:2019vlb,Vagnozzi:2020gtf}, as well as the \textit{Planck} 2018 constraints on inflation paper~\cite{Planck:2018jri}, for an inevitably incomplete example of recent works adopting this assumption. Other commonly adopted simplifying assumptions include that of a standard thermal history with an instantaneous transition from the early-Universe phase responsible for sourcing the primordial fluctuations (be it inflation or an alternative mechanism, usually taking place at a very high energy scale) to the phase of radiation domination, i.e. an instantaneous reheating phase with no extended matter or stiff matter domination era prior to the usual radiation domination era.

However, all of these are clearly approximations at best, at least for the case of the inflationary SGWB. For instance, by making use of the powerful inflationary flow formalism, it was recently argued in Ref.~\cite{Kinney:2021nje} that in an ensemble of realistic single-field inflationary models, a power-law extrapolation from CMB to interferometer scales can overestimate the amplitude of primordial tensor modes by up to two orders of magnitude. The largest frequencies are sensitive to the non-slow-roll dynamics towards the end of inflation, where the expansion rate and slow roll parameters vary relatively rapidly, typically leading to one or more breaks in the primordial tensor spectrum (see e.g. Refs.~\cite{Kinney:2005in,Caligiuri:2014ola,Giare:2020vhn,Vagnozzi:2020gtf} for related discussions). As a result, the tensor spectral index measured on CMB scales can be completely uncorrelated from the same quantity probed on interferometer scales. We remark, however, that the analysis of Ref.~\cite{Kinney:2021nje} strictly speaking only applies to single-field models which, as discussed in the Introduction, we know cannot accommodate the NANOGrav signal due to the latter requiring a strongly blue GW spectrum.

\subsection{Broken power-law spectrum}
\label{subsec:brokenpowerlaw}

To consider a more realistic SGWB spectrum, we go beyond the simple assumption of a pure power-law across various decades in frequency discussed previously. In particular, we consider a phenomenological broken power-law scenario (which however we shall later justify on a theoretical basis) effectively corresponding to the SGWB power-law index changing from $n_T$ to $\alpha$ at a characteristic scale $k_{\alpha}$~\cite{Kuroyanagi:2014nba}. More specifically, the dimensionless GW density parameter $\Omega_{\rm GW}^{\rm bpl}$ (with ``bpl'' standing for ``broken power-law'') is related to the previously introduced $\Omega_{\rm GW}^{\rm pl}$ [itself associated to the pure power-law primordial tensor spectrum given in Eq.~(\ref{eq:ptpowerlaw})] as follows:
\begin{eqnarray}
\Omega_{\rm GW}^{\rm bpl} = 
\begin{cases}
\Omega_{\rm GW}^{\rm pl} & (k<k_{\alpha})\,, \\
\Omega_{\rm GW}^{\rm pl} \left ( \frac{k}{k_{\alpha}} \right ) ^{\alpha} & (k>k_{\alpha})\,.
\end{cases}
\label{eq:omegagwbrokenpowerlaw}
\end{eqnarray}
where the scale $k_{\alpha}$ scale corresponds to a break frequency $f_{\alpha} \sim 1.5 \times 10^{-15}(k_{\alpha}/{\rm Mpc}^{-1})\,{\rm Hz}$. We envisage the break frequency to be ${\cal O}({\rm nHz}) \lesssim f_{\alpha} \lesssim {\cal O}({\rm Hz}),$ so that the NANOGrav signal falls within the first half of the spectrum, and constraints from LIGO/Virgo are relevant to the post-break spectrum. On the other hand, the break can allow us to fit the NANOGrav signal while not running afoul of constraints on the SGWB amplitude on ${\cal O}(10)\,{\rm Hz}$ frequencies from LIGO/Virgo, provided $\alpha<0$ (or at the very least $\alpha<n_T$), so that the post-break spectrum is less strongly blue, if not red. In our analysis, we shall require $\alpha<0$, so that the high-frequency part of the SGWB spectrum is red, as this is the only way a possible SGWB detection from NANOGrav can be reconciled with LIGO/Virgo's non-detection.

While the choice of introducing the broken power-law parametrization in Eq.~(\ref{eq:omegagwbrokenpowerlaw}) is phenomenological in nature, we shall now argue that it is theoretically justified, as it can actually cover a very broad range of early Universe scenarios (as discussed in more detail in Ref.~\cite{Kuroyanagi:2014nba}). It is important to note that, in Eq.~(\ref{eq:omegagwbrokenpowerlaw}) we have introduced the broken power-law parametrization at the level of dimensionless GW density today. This can be interpreted as arising from at least two fundamentally different classes of scenarios: a first one where it is the transfer function which is modified (for instance due to the propagation of primordial GWs through non-standard epochs of the expansion of the Universe) leading to the broken power-law, and a second one where it is instead the primordial tensor power spectrum which features a broken power-law, where the break could be due to processes inherent to the GW production mechanism. At the level of $\Omega_{\rm GW}^{\rm bpl}$, these two distinct theoretical origins are completely degenerate. While we shall discuss in detail various theoretical scenarios leading to a break in the GW spectrum, and falling within either of the two categories described above, in our later analysis we shall remain completely agnostic as to the underlying theoretical origin of the break, while simply adopting the phenomenological view that Eq.~(\ref{eq:omegagwbrokenpowerlaw}) can cover a broad range of interesting early Universe scenarios.

The broken power-law spectrum is best suited to describe models with a non-standard background evolution, \textit{i.e.} where the Universe does not transition from inflation to radiation domination following an instantaneous reheating process~\cite{Kuroyanagi:2014nba,Cook:2015vqa,Kuroyanagi:2020sfw}. For example, an extended non-instantaneous reheating period, possibly with a low reheating scale, leads to an extended early matter domination before the usual radiation domination: this scenario leads to a break in the GW density which is captured by setting $\alpha=n_T-2$, with $k_{\alpha}$ determined by the reheating temperature.~\footnote{In this case the exact relation between the break wavenumber and the reheating temperature is given by Eq.~(14) in Ref.~\cite{Kuroyanagi:2014nba}.}

More generally, if the Universe is dominated by a fluid with effective equation of state $w_{\rm eff}$ prior to radiation domination, the effect is also that of a break in the GW power spectrum, captured by setting $\alpha=n_T+2(3w_{\rm eff}-1)/(1+3w_{\rm eff})$, with $f_{\alpha}$ related to the temperature at which the Universe switches to being radiation dominated. In particular, this expression recovers the case of an extended matter domination era, where $w_{\rm eff}=0$ and therefore $\alpha=n_T-2$. Another theoretically interesting case is that where the Universe undergoes a kination or stiff matter domination phase, where $w_{\rm eff}=1$ and therefore $\alpha=n_T+1$ (see e.g. Ref.~\cite{Li:2016mmc}). However, this case will not be of interest to our work, as it makes the GW spectrum even bluer on small scales, thereby worsening the disagreement with LIGO/Virgo. The broken power-law spectrum can also describe the effect of late-time entropy injection (for instance through the decay of an additional scalar field other than the inflaton), although in this case the exact value of $\alpha$ depends on the specific entropy injection details~\cite{Kuroyanagi:2014nba}.

The scenarios described above effectively lead to a change in the transfer function appearing in Eq.~(\ref{eq:ptpowerlaw}), which generally takes a factorizable form with the various factors accounting for different physical processes (e.g. the standard radiation-to-matter transition, or the non-standard scenarios described above). However, as alluded to earlier, the broken power-law spectrum can also effectively describe scenarios where it is the underlying primordial power spectrum which features a break in the spectral index $n_T$, which is therefore non-uniform across the frequency range. In fact, many early Universe models predicting a blue spectrum on CMB scales typically predict a break in the spectral index on smaller scales. One class of examples are inflationary models where the inflaton is coupled to the field-strength tensor of vector (gauge) fields (see e.g. Refs.~\cite{Dimastrogiovanni:2018xnn,Maleknejad:2018nxz,Maleknejad:2019hdr,Wolfson:2020fqz,Wolfson:2021fya}). In these models, production of gauge fields is typically effective over an extended period during inflation, leading to a blue spectrum for modes which exit the horizon during this period (see e.g.~\cite{Cook:2011hg,Pajer:2013fsa,Mukohyama:2014gba}). However, when gauge field production eventually stops being efficient, the associated spectrum returns to being red, decreasing at higher frequencies. Alternatives to inflation may also predict a break in the spectral index of the primordial tensor spectrum.

In this work, we shall therefore adopt $\Omega_{\rm GW}^{\rm bpl}$ as a phenomenological but well-motivated choice beyond the pure power-law approximation utilized in many works. Given our choice of setting $\alpha<0$ so that the high-frequency part of the SGWB spectrum is red, the choice of $f_{\rm UV}$ ends up playing a very marginal role. In fact, since the SGWB spectrum is blue for $f<f_{\alpha}$ and red for $f>f_{\alpha}$, integrated quantities such as $\rho_{\rm GW}$ or equivalently $N_{\rm eff}$ are mostly affected by the form of the spectrum around $f_{\alpha}$: the low-frequency part of the integral is mostly insensitive to $f_{\rm IR}$ being the respective part of the spectrum blue, and similarly for the high-frequency part of the integral being mostly insensitive to $f_{\rm UV}$. For this reason, in the following, we shall adopt the conservative choice of setting $f_{\rm UV} \sim 10^8\,{\rm Hz}$. As discussed earlier in Sec.~\ref{subsec:power-law}, this choice arises by requiring that the power-law spectrum extends over $\sim$60 \textit{e}-folds~\cite{Meerburg:2015zua,Cabass:2015jwe}, while not committing to any specific (inflationary or non) early Universe scenario. However, the less conservative choice of $f_{\rm UV} \sim 10^{43}\,{\rm Hz}$ would not qualitatively affect our results.

Before closing, a comment on some of the non-minimal primordial scenarios discussed above is in order: in particular, we have mentioned the possibility of low-scale reheating and/or a non-standard post-reheating background expansion such as kination, which modify the expansion of the Universe prior to the usual radiation domination phase. While non-minimal, these scenarios are far from being exotic, and are not hard to come by under rather generic assumptions. For examples, scenarios where reheating occurs at lower temperatures or is delayed (in some cases due to known Standard Model physics such as the dynamics of the Higgs boson) have been studied in e.g. Refs.~\cite{Kawasaki:2000en,Giudice:2000ex,Hannestad:2004px,Litsa:2020rsm}, with the recent analysis of Ref.~\cite{deSalas:2015glj} finding that reheating temperatures as low as $T_{\rm rh} \sim {\cal O}({\rm MeV})$ are consistent with \textit{Planck} data. On the other hand, in kination scenarios the expansion rate of the post-inflationary Universe is dominated by the kinetic energy of a fast-rolling scalar field, leading to an effective equation of state $w_{\rm eff} \sim 1$~\cite{Ford:1987de}. Such scenarios have been studied in detail in recent years, particularly in light of their possible implications for the production of dark matter~\cite{Visinelli:2015eka,Redmond:2017tja,Visinelli:2017qga,Visinelli:2018wza}.

The assumption of a break in the SGWB spectrum makes it particularly important to constrain such a spectrum using a multi-frequency approach. That is, making use of observations sensitive to the GW energy density across as wide a frequency range as possible. In the following, we shall therefore discuss the diverse class of cosmological and astrophysical observations we shall use to constrain this scenario.

\section{Constraints on gravitational waves across the frequency spectrum}
\label{sec:sources}

The landscape of current and future probes of GWs is extremely vast, diverse, and complementary, allowing to probe the signatures of GWs across a wide range of times and frequencies. Some of these probes (\textit{e.g.} interferometers) are sensitive to the SGWB in a certain relatively narrow frequency band, whereas other probes carry an integral sensitivity to the SGWB energy density in a wide frequency range (\textit{e.g.} BBN). Below we briefly present the GW probes we will make use of in this work.

\begin{figure*}[t]
\centering
\includegraphics[width=0.6\linewidth]{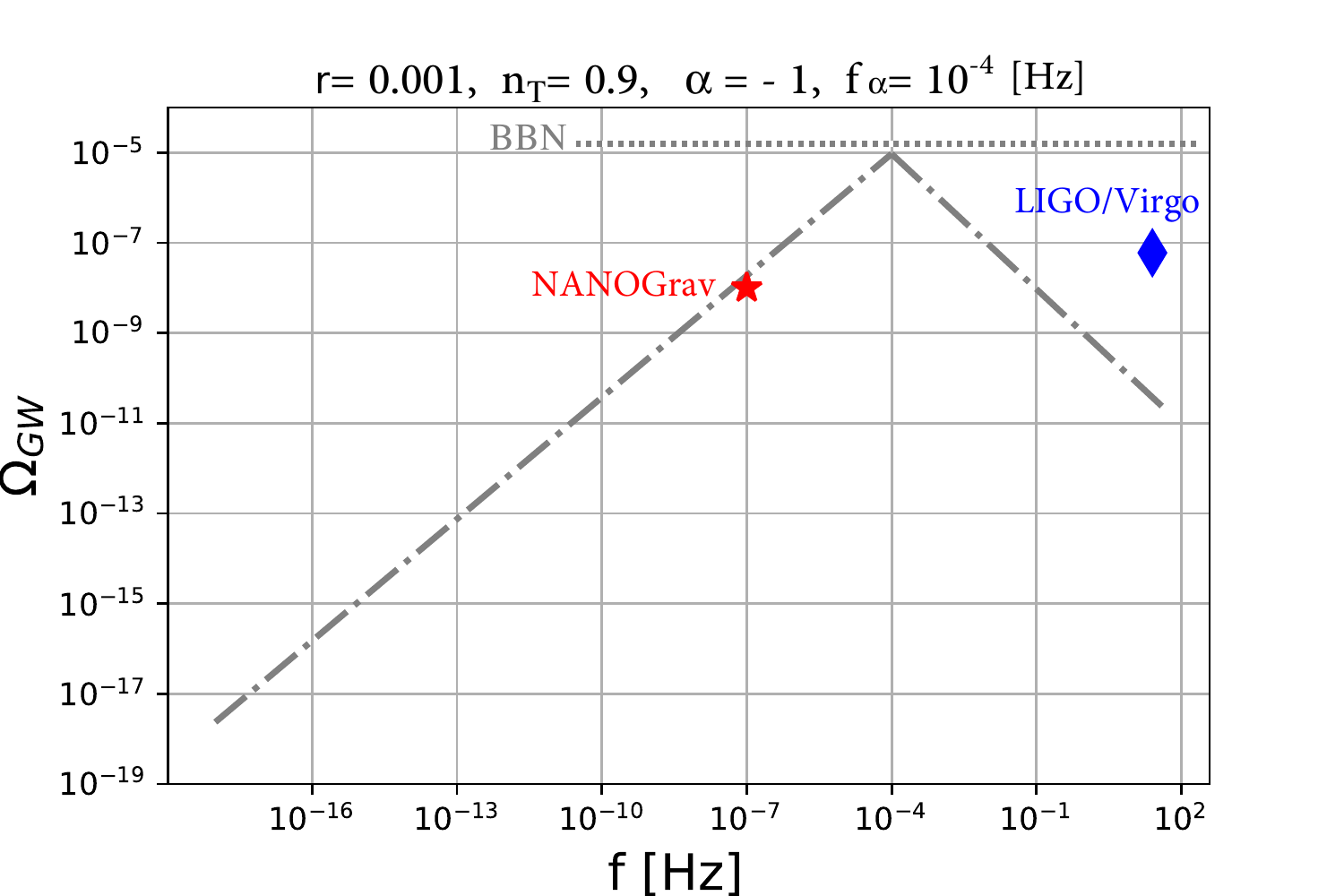}
\caption{Benchmark example (consistent with data) of the broken power-law SGWB spectrum considered in this work (grey dashed-dotted line), with tensor-to-scalar ratio $r=10^{-3}$, pre-break tensor spectral index $n_T=0.9$, break frequency $f_{\alpha}=10^{-4}\,{\rm Hz}$, and post-break tensor spectral index $\alpha=-1$. The plot indicates the tentative NANOGrav signal (red star) and LIGO/Virgo's upper limit (blue diamond, where O1 and O3 stands for first and third observing run respectively), as well as an indicative BBN limit on the SGWB energy density (grey dotted line): as is clearly seen seen, a break in the GW spectrum is required to reconcile a blue spectrum explaining NANOGrav with LIGO/Virgo's upper limit. This figure is based on Fig.~1 of Ref.~\cite{Kuroyanagi:2020sfw}, in order to provide a direct parallel with their results.}
\label{fig:OmegaGW_vs_experiments}
\end{figure*}

\subsection{Interferometers}
\label{subsec:gravitationalwaveobservatories}

Large laser interferometers can be used for GW direct detection, through the effect of a passing GW shortening or lengthening the two arms of the interferometers, in turn affecting the resulting interferometric patterns. Two state-of-the-art current GW interferometers are the Laser Interferometer Gravitational-Wave Observatory (LIGO) and the Virgo interferometer~\cite{VIRGO:2014yos,LIGOScientific:2014pky}, both of which are sensitive to GWs in the $10$-$10^4\,{\rm Hz}$ frequency range. To date, the LIGO and Virgo collaborations have detected $\sim$ 90 resolved GW events resulting from the mergers of stellar objects (black holes and neutron stars) \cite{LIGOScientific:2021djp}, and in doing so opened a remarkable window onto the Universe and revolutionized our understanding of gravity, in part thanks to coincidental multi-messenger observations across the electromagnetic spectrum~\cite{Yunes:2016jcc,Creminelli:2017sry,Sakstein:2017xjx,Ezquiaga:2017ekz,Baker:2017hug,Boran:2017rdn,Visinelli:2017bny,Langlois:2017dyl,Cai:2018rzd,Pardo:2018ipy,Casalino:2018tcd,Casalino:2018wnc,Frusciante:2020gkx,Sakstein:2020axg,DeLuca:2020sae}.

Besides resolved events, interferometers can be used to search for a possible astrophysical or cosmological SGWB, such as the one we are considering in this work. LIGO and Virgo have placed an upper limit on the amplitude of the SGWB in the frequency range $20 \lesssim f/{\rm Hz} \lesssim 86$. Following other works, we shall take the following 95\%~confidence level (C.L.) upper limit~\cite{LIGOScientific:2016jlg,LIGOScientific:2019vic}:~\footnote{Strictly speaking, this limit assumes a fiducial value for $H_0$ (which should scale in such a way that $\Omega_{\rm GW} \propto H_0^{-2}$ when scanning over $H_0$ in our subsequent analysis). However, we do not expect this point to have a significant impact on our results.}
\begin{eqnarray}
\Omega_{\rm GW}(k_{\rm LV}) \lesssim 1.7 \times 10^{-7}\,,
\label{eq:ligolimit}
\end{eqnarray}
where $k_{\rm LV} \sim 2.3 \times 10^{16}\,{\rm Mpc}^{-1}$ is the comoving wavenumber corresponding to a frequency $f_{\rm LV} \sim 35\,{\rm Hz}$ which we take as representative for LIGO/Virgo's limit (following earlier work), and ``LV'' stands for ``LIGO/Virgo''. More recent constraints place an even tighter limit, $\Omega_{\rm GW}(k_{\rm LV}) \lesssim 6.6 \times 10^{-9}$, at a frequency of $f \sim 25\,{\rm Hz}$~\cite{KAGRA:2021kbb}.~\footnote{The data products of LIGO/Virgo O3 are available at: \url{https://dcc.ligo.org/LIGO-G2001287/public}}.

We note that the limit in Eq.~(\ref{eq:ligolimit}) is sensitive to the SGWB energy density at a specific wavenumber. In particular, the frequency $f_{\rm LV} \sim 35\,{\rm Hz}$ falls within the post-break part of the broken power-law GW spectrum of Eq.~(\ref{eq:omegagwbrokenpowerlaw}). Therefore only the part of the spectrum for $k>k_{\alpha}$, or equivalently $f>f_{\alpha}$, needs to be considered in order for theoretical predictions to be compared to LIGO/Virgo's upper limit in Eq.~(\ref{eq:ligolimit}). Hence, we expect LIGO/Virgo constraints on the SGWB amplitude to mostly constrain the break frequency $f_{\alpha}$ and post-break spectral index $\alpha$. For purely illustrative purposes, in Fig.~\ref{fig:OmegaGW_vs_experiments} we show a benchmark example of the broken power-law SGWB spectrum, indicating the NANOGrav signal and LIGO/Virgo's upper limit. As can be seen, a break in the GW spectrum is clearly required to reconcile a pre-break blue spectrum explaining NANOGrav with LIGO/Virgo's upper limit.

Looking to the near future, prospects for direct detection of resolved GW events and the SGWB with interferometers and other types of surveys are very bright. The frequency window between ${\cal O}(10^{-7})$ and ${\cal O}(10^3)\,{\rm Hz}$ will be covered by a diverse range of experiments, including but not limited to space-based laser interferometers, next-generation ground-based detectors, binary resonance probes, and space-based atom interferometry. In Sec.~\ref{sec:conclusions} we will discuss in more detail future probes of GWs (including the broken power-law SGWB spectrum we are considering) across the GW frequency landscape.

\subsection{Pulsar Timing Arrays}
\label{subsec:pulsartimingarrays}

Pulsar timing arrays (PTAs) aim for a SGWB detection by exploiting the fact that millisecond pulsars behave as extremely stable clocks. A SGWB travelling between an ensemble of pulsars and us would leave its imprint through fluctuations in the arrival times of radio pulses, which would be spatially correlated~\cite{1978SvA....22...36S,Detweiler:1979wn,1990ApJ...361..300F}. By searching for these correlations, PTAs can search for a SGWB in the $10^{-9}-10^{-7}\,{\rm Hz}$ frequency range. Current and planned PTA surveys include NANOGrav~\cite{NANOGrav:2020bcs}, PPTA~\cite{Kerr:2020qdo}, and EPTA~\cite{Lentati:2015qwp}, collectively constituting the IPTA~\cite{Perera:2019sca}. Moreover, the Square Kilometer Array is expected to potentially be able to detect thousands of millisecond pulsars, and thus might play an important role in the landscape of future PTA surveys~\cite{Weltman:2018zrl}.

It is customary to report the results of PTA searches in terms of the GW strain power spectrum as a function of frequency, $h_c(f)$, which is related to $\Omega_{\rm GW}(f)$ by:
\begin{eqnarray}
\Omega_{\rm GW}(f) =  \frac{2\pi^2}{3H_0^2}f^2h_c^2(f)\,.
\label{PA1}
\end{eqnarray}
The GW strain power spectrum is typically approximated as a power-law at a reference frequency $f_{\rm yr}=1\,{\rm yr}^{-1}$ (a numerically convenient value given the frequencies to which PTAs are most sensitive), with amplitude and spectral index given by $A_{\rm CP}$ and $\alpha_{_{\rm CP}}$ respectively:
\begin{eqnarray}
h_c(f)=A_{\rm CP} \left ( \frac{f}{f_{\rm yr}} \right ) ^{\alpha_{_{\rm CP}}} \equiv A_{\rm CP} \left ( \frac{f}{f_{\rm yr}} \right ) ^{\frac{3-\gamma_{_{\rm CP}}}{2}} \,,
\label{PA2}
\end{eqnarray}
where the spectral index $\alpha_{_{\rm CP}}$ is often exchanged for the related quantity $\gamma_{_{\rm CP}} \equiv 3-2\alpha_{_{\rm CP}}$. The SGWB resulting from merging SMBHBs is expected to be described by $\gamma_{_{\rm CP}}=13/3$ or, equivalently, $\alpha_{_{\rm CP}}=-2/3$~\cite{Rajagopal:1994zj,Wyithe:2002ep,Sesana:2013wja}.

In this work we shall consider the NANOGrav signal~\cite{NANOGrav:2020bcs}, which with all the caveats discussed earlier we shall interpret as a genuine SGWB detection. The NANOGrav collaboration fitted the power-law approximation of the strain power spectrum in Eq.~(\ref{PA2}) to their 5 lowest frequency bins with highest signal-to-noise, in the frequency range $2.5 \times 10^{-9} \lesssim f/{\rm Hz} \lesssim 9.0 \times 10^{-8}$, and obtained joint constraint on $\log_{10}A_{_{\rm CP}}$ and $\gamma_{_{\rm CP}}$. We note that, for a SGWB spectrum behaving as a pure power-law at least up to the NANOGrav frequency range, such as the one we are considering since the break occurs for $f_{\alpha} \gtrsim {\cal O}({\rm nHz})$, the tensor spectral index $n_T$ is related to $\gamma_{_{\rm CP}}$ and/or $\alpha_{_{\rm CP}}$ as $n_T=5-\gamma_{_{\rm CP}}=2+2\alpha_{_{\rm CP}}$, whereas $A_{_{\rm CP}}$ scales as $A_{_{\rm CP}} \propto \sqrt{r}$~\cite{Zhao:2013bba,Vagnozzi:2020gtf}. 


\begin{figure*} [t]
\centering
\includegraphics[width=0.45\linewidth]{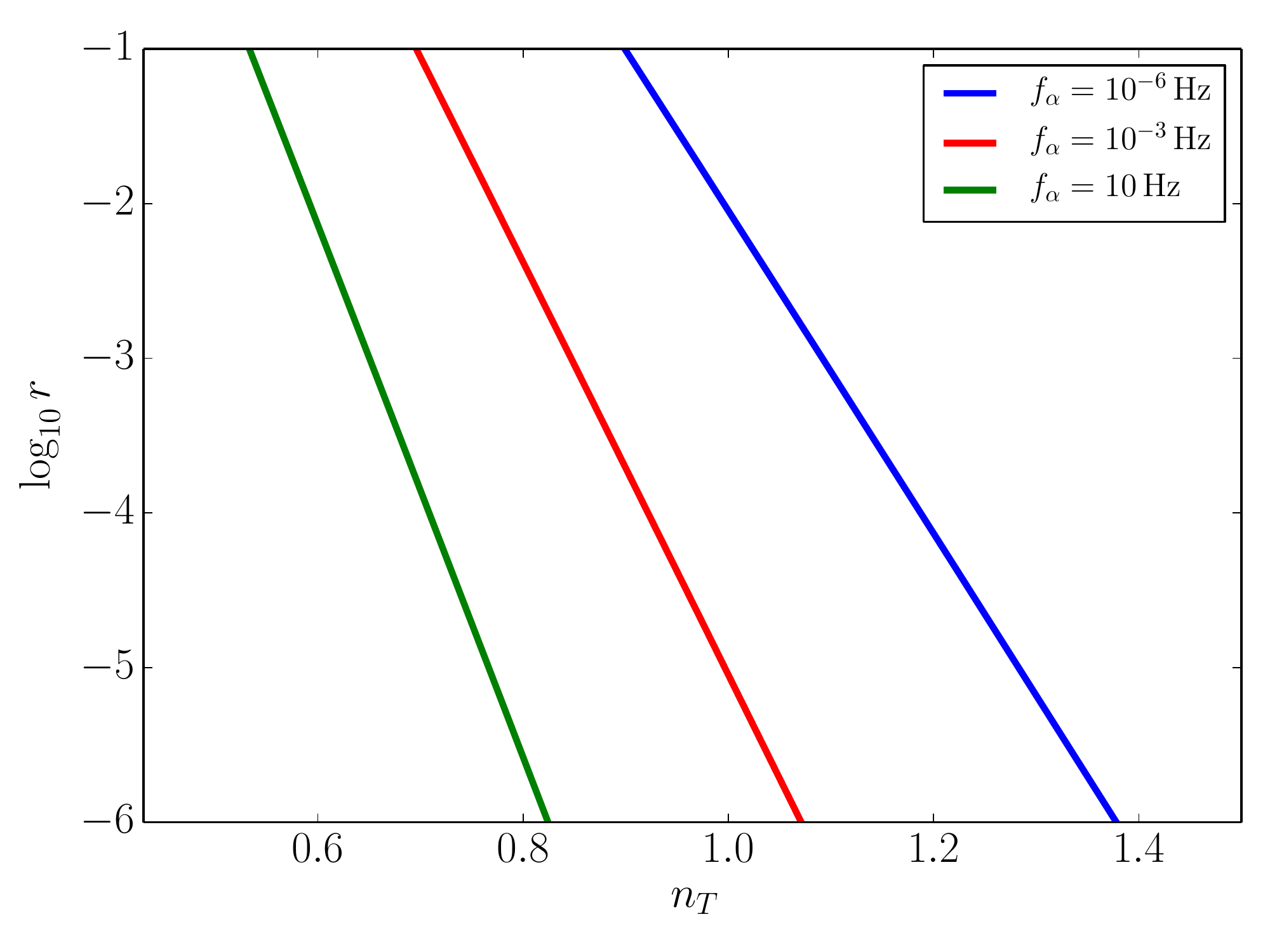}
\includegraphics[width=0.45\linewidth]{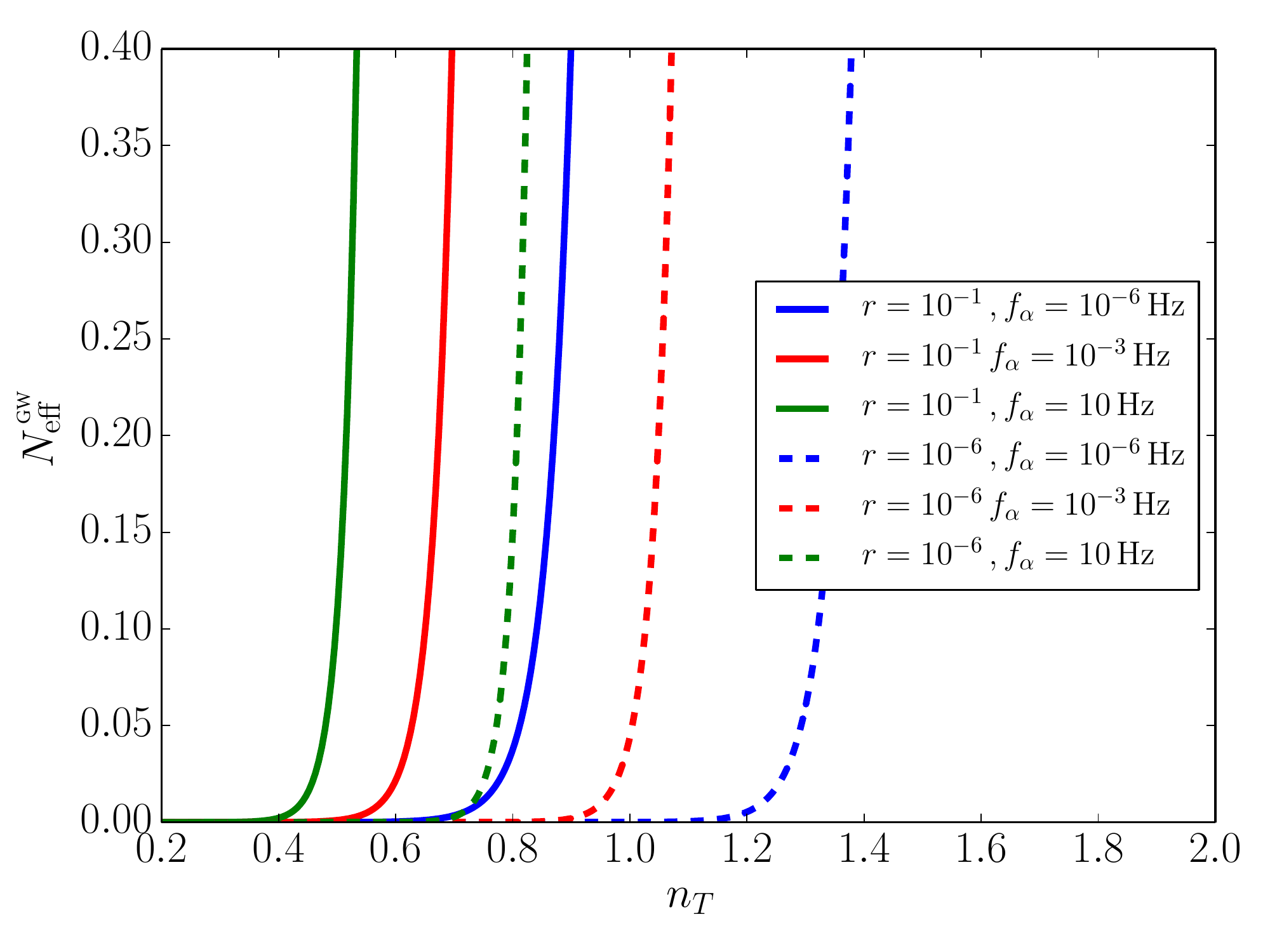}
\caption{Evolution of $N_{\rm eff,BBN}^{\rm GW}$ as a function of $n_T$, $r$, and $f_{\alpha}$, as captured by Eq.~(\ref{NeffBBNapprox}), neglecting the $-1/\alpha$ term which is subdominant as argued in the main text. \textit{Left panel}: curves of constant $N_{\rm eff,BBN}^{\rm GW}$ in the $\log_{10}r$-$n_T$ plane, saturating the $N_{\rm eff,BBN}^{\rm GW} \lesssim 0.4$ limit, indicatively corresponding to the upper limit on the amount of extra radiation set by BBN, for different values of the break frequency $f_{\alpha}$ (as per the color coding). \textit{Right panel}: evolution of $N_{\rm eff,BBN}^{\rm GW}$ as a function of $n_T$ for different values of the tensor-to-scalar ratio $r$ and break frequency $f_{\alpha}$ (as per the color coding).}
\label{fig:eq15}
\end{figure*}

\subsection{Big Bang Nucleosynthesis}
\label{subsec:bbn}

Big Bang Nucleosynthesis (BBN), taking place in the very earliest stages of our universe, is the process responsible for the production of light nuclei other than those of $^1$H (see e.g. Ref.~\cite{Cyburt:2015mya} for a review). BBN is primarily responsible for the production of $^4$He, alongside smaller amounts of $^3$He, deuterium ($^2$H) and tritium ($^3$H), $^7$Li, and $^7$Be, with $^3$H and $^7$Be later decaying to $^3$He and $^7$Li respectively. The final yield of light elements, tightly constrained observationally, is highly sensitive to the expansion rate of the Universe, usually assumed to proceed as in a radiation-dominated Universe following reheating in the standard scenario. Therefore, BBN will be highly sensitive to a non-standard SGWB spectrum such as the one considered, given that blue primordial GWs contribute to the energy density of the Universe as an extra radiation component.

Unlike the previously discussed constraints from interferometers (LIGO/Virgo) and PTAs (NANOGrav), which are mostly sensitive to the SGWB spectral density at a certain frequency ($f_{\rm LV}$ and $f_{\rm yr}$ respectively) or within a narrow frequency band, BBN carries integrated sensitivity to a wide range of frequencies from $f_{\rm IR,BBN}$ to $f_{\rm UV}$, as captured by an effective number of relativistic species $N_{\rm eff,BBN}^{\rm GW}$ defined analogously to Eq.~(\ref{eq:rhotot}) starting from Eq.~(\ref{eq:rhogw}). In this case, the relevant IR cutoff is given by $f_{\rm IR,BBN} \sim 10^{-10}\,{\rm Hz}$, approximately corresponding to the comoving horizon at the time of BBN, when the temperature of the Universe was $T \sim {\cal O}({\rm MeV})$. For the broken power-law model, it is relatively straightforward to show that $N_{\rm eff,BBN}^{\rm GW}$ is approximately given by:
\begin{align}
N_{\rm eff,BBN}^{\rm GW} \sim &0.3\frac{rA_s}{n_T} \left [ \left ( \frac{f}{f_{\star}} \right ) ^{n_T} \right ] ^{f_{\alpha}}_{f_{\rm IR,BBN}} \nonumber \\
+&0.3\frac{rA_s}{\alpha} \left [ \left ( \frac{f}{f_{\alpha}} \right ) ^{\alpha} \right ] ^{f_{\rm UV}}_{f_{\alpha}}\,,
\label{NeffBBN}
\end{align}
%
%
with pivot scale given by $k_{\star}=0.01\,{\rm Mpc}^{-1}$ (or equivalently $f_{\star}=1.5 \times 10^{-17}\,{\rm Hz}$), and as stated earlier we fix $f_{\rm IR,BBN}= 10^{-10}\,{\rm Hz}$ and $f_{\rm UV}=10^8\,{\rm Hz}$~\cite{Meerburg:2015zua,Cabass:2015jwe}. Inserting numbers, the above expression becomes:
\begin{align}
N_{\rm eff,BBN}^{\rm GW} \sim &0.3\frac{rA_s}{n_T} \left [ \left ( \frac{2}{3}10^{17} \left ( \frac{f_{\alpha}}{{\rm Hz}} \right ) \right ) ^{n_{T}} - \left ( \frac{2}{3} 10^{7} \right ) ^{n_T} \right ] \nonumber \\
+&0.3\frac{rA_s}{\alpha} \left [ \left ( 10^{-8} \left ( \frac{f_{\alpha}}{{\rm Hz}} \right ) \right ) ^{\vert \alpha \vert} - 1 \right ] \,,
\label{NeffBBNfinal}
\end{align}
where $\vert \alpha \vert = -\alpha$, since we require $\alpha<0$ in order for the high-frequency part of the SGWB spectrum to be red and satisfy LIGO/Virgo limits, while simultaneously explaining the NANOGrav signal with the blue low-frequency part of the SGWB spectrum.

Let us examine more closely the two contributions to $N_{\rm eff,BBN}^{\rm GW}$ in Eq.~(\ref{NeffBBNfinal}), with the first line corresponding to the low-frequency part of the spectrum ($f<f_{\alpha}$), and the second line corresponding to the high-frequency one ($f>f_{\alpha}$). Unsurprisingly, we see that the contribution to $N_{\rm eff,BBN}^{\rm GW}$ from the low-frequency part of the broken power-law spectrum is dominated by the highest available frequencies (\textit{i.e.} those around the break $f_{\alpha}$), and similarly the contribution to $N_{\rm eff,BBN}^{\rm GW}$ from the high-frequency part is dominated by the lowest available frequencies (\textit{i.e.} again those around the break $f_{\alpha}$). As anticipated earlier in Sec.~\ref{subsec:brokenpowerlaw}, this occurs as we are within the regime where $n_T>0$ and $\alpha<0$, \textit{i.e.} where the low-/high-frequency part of the SGWB spectrum is respectively blue/red. Therefore, the blue (low-frequency) part of the spectrum will be most sensitive to the highest available frequencies, and conversely for the red (high-frequency) part, implying that $N_{\rm eff,BBN}^{\rm GW}$ ends up being dominated by modes around $f_{\alpha}$.

Neglecting sub-dominant terms, we find that Eq.~(\ref{NeffBBNfinal}) can be well approximated by:
\begin{eqnarray}
N_{\rm eff,BBN}^{\rm GW} \approx \frac{0.3rA_s}{n_T} \left [ \left ( \frac{2}{3}10^{17} \left ( \frac{f_{\alpha}}{{\rm Hz}} \right ) \right ) ^{n_T}-\frac{n_T}{\alpha} \right ] \,.
\label{NeffBBNapprox}
\end{eqnarray}
Observational determinations of the abundances of light elements severely restrict $N_{\rm eff,BBN}^{\rm GW}$. Different probes return different limits, but we can safely take $N_{\rm eff,BBN}^{\rm GW} \lesssim 0.4$ as an indicative upper limit~\cite{Aver:2015iza,Cooke:2017cwo,Planck:2018vyg,Hsyu:2020uqb,Mossa:2020gjc}. Moreover, we find that for $f_{\alpha}$ within the range $10^{-9}\,{\rm Hz} \lesssim f_{\alpha} \lesssim 35\,{\rm Hz}$, \textit{i.e.} within the NANOGrav and LIGO/Virgo frequencies, the $-n_T/\alpha$ term is always several orders of magnitude smaller than the $f_{\alpha}$-dependent term, and hence can be safely neglected.~\footnote{To see this, one can estimate the last term on the right-hand side of Eq.~(\ref{NeffBBNapprox}) with the aid of Eq.~(\ref{eq:omegagwbrokenpowerlaw}). A lower limit to $\vert \alpha \vert$ can be obtained by considering the lowest possible value of $f_{\alpha}$, corresponding to the NANOGrav frequency $f \sim 3 \times 10^{-8}\,{\rm Hz}$, and saturating the inequality set by the LIGO/Virgo limit. Setting the left-hand side of Eq.~(\ref{eq:omegagwbrokenpowerlaw}) to the LIGO/Virgo upper limit ($\Omega_{\rm GW} \sim 1.7 \times 10^{-7}$), relating this to the SGWB energy density $\Omega_{\rm GW}$ indicated by NANOGrav ($\Omega_{\rm GW} \sim 5.6 \times 10^{-9}$), and using the fact that $k/k_{\alpha}=f_{\rm LV}/f_{\alpha}= 35/(3 \times 10^{-8})$, we find the rough lower limit $\vert \alpha \vert \gtrsim 0.15$. It is trivial to show that for any value of the tensor spectral index $0.7 \lesssim n_T \lesssim 1.3$ (required to explain NANOGrav, as shown in Ref.~\cite{Vagnozzi:2020gtf}), the last term on the right-hand side of Eq.~(\ref{NeffBBNapprox}) is always at least 17 orders of magnitude smaller than the first term.} In any case we stress that, within our subsequent numerical study, we use the full expression for $N_{\rm eff,BBN}^{\rm GW}$ given in Eq.~(\ref{NeffBBN}).

The evolution of $N_{\rm eff,BBN}^{\rm GW}$ as a function of $n_T$, $r$, and $f_{\alpha}$ is shown in Fig.~\ref{fig:eq15}. In particular, the left panel shows curves of constant $N_{\rm eff,BBN}^{\rm GW}$ in the $\log_{10}r$-$n_T$ plane saturating the $N_{\rm eff,BBN}^{\rm GW} \lesssim 0.4$ limit, whereas the right panel shows the evolution $N_{\rm eff,BBN}^{\rm GW}$ as a function of $n_T$ for different values of $r$ and $f_{\alpha}$. In producing Fig.~\ref{fig:eq15}, we neglect the $\alpha$-dependent term in Eq.~(\ref{NeffBBNapprox}) as argued earlier. From the left panel, we see that the choice of $f_{\alpha}$ can significantly influence the allowed value of $n_T$: the lower the break frequency, the larger $n_T$ is allowed to be. This is again unsurprising: reducing $f_{\alpha}$ reduces the contribution to $N_{\rm eff,BBN}^{\rm GW}$ from the highest available frequencies in the low-frequency part of the SGWB spectrum, as is clear from the first line of Eq.~(\ref{NeffBBNfinal}). This behavior, wherein lower values of $f_{\alpha}$ allow for higher values of $n_T$, is reminiscent of a similar behavior observed by one of us in Ref.~\cite{Vagnozzi:2020gtf} when considering different reheating temperatures $T_{\rm rh}$, as there is approximately a one-to-one correspondence between $f_{\alpha}$ and $T_{\rm rh}$: for instance reheating temperatures $T_{\rm rh} \sim 10^{10}/10^5/100\,{\rm GeV}$ approximately correspond to break frequencies $f_{\alpha} \sim 100/10^{-3}/10^{-6}\,{\rm Hz}$ respectively. From the right panel we instead see that the higher the tensor-to-scalar ratio $r$ and break frequency $f_{\alpha}$, the more rapidly $N_{\rm eff,BBN}^{\rm GW}$ increases with $n_T$, a fact which is again unsurprising.

\subsection{Cosmic Microwave Background}
\label{subsec:cmb}

The Cosmic Microwave Background (CMB), the relic radiation from the epoch of recombination, is an extremely important probe of early-Universe physics. For what concerns our study, we will be interested in the CMB at two different levels. Firstly, the CMB probes GWs at extremely low frequencies ($f \lesssim 10^{-16}\,{\rm Hz}$), through their imprint on the B-mode polarization pattern: therefore, measurements of the B-mode power spectrum, and in particular limits on the tensor-to-scalar ratio $r$ (e.g. Ref.~\cite{Planck:2018vyg}), can directly constrain the low-frequency part of our SGWB spectrum.

Secondly, the CMB is highly sensitive to the energy density of any extra radiation component, particularly through the effect of the latter on the damping of tail of the small-scale (large multipole $\ell$) power spectra (Silk damping)~\cite{Silk:1967kq} and on the early integrated Sachs-Wolfe (eISW) effect, both of which are tightly constrained by \textit{Planck} CMB data~\cite{Hou:2011ec,Cabass:2015xfa,Kable:2020hcw,Vagnozzi:2021gjh}. As with BBN, the CMB is sensitive to a large range of modes, from $f_{\rm IR,CMB}$ to $f_{\rm UV}$. Therefore, the expression for $N_{\rm eff,CMB}^{\rm GW}$ ends up being analogous to the BBN one, given in Eq.~(\ref{NeffBBN}), except for the lower integration limit in the first line (low-frequency contribution) being given by $f_{\rm IR,CMB}$ instead of $f_{\rm IR,BBN}$.

In principle not only is $f_{\rm IR,CMB}$ much lower than $f_{\rm IR,BBN} \sim 10^{-10}\,{\rm Hz}$, but it is a time-dependent quantity, as at any given time only sub-horizon modes oscillate and propagate as extra radiation components. However, to simplify the discussion, we can again resort to the fact that the low-frequency part of the SGWB spectrum is blue. This leads to the expectation that the most important low-frequency contributions to $N_{\rm eff,CMB}^{\rm GW}$ come from modes around the break frequency $f_{\alpha}$, and similarly for the high-frequency contributions, given that the high-frequency part of the SGWB is red. In practice, this implies that $N_{\rm eff,CMB}^{\rm GW}$ is mostly insensitive to $f_{\rm IR,CMB}$ and $f_{\rm UV}$, given that $f_{\rm IR,CMB} \ll 10^{-10}\,{\rm Hz} \ll f_{\alpha}$ and $f_{\rm UV} \sim 10^8\,{\rm Hz} \gg f_{\alpha}$. Formally, this means that we can safely set $f_{\rm IR,CMB} \sim 0\,{\rm Hz}$, and that the expression for $N_{\rm eff,CMB}^{\rm GW}$ (relevant for the computation of the CMB temperature and E-mode polarization anisotropy power spectra) is identical to that for $N_{\rm eff,BBN}^{\rm GW}$ in Eq.~(\ref{NeffBBN}).

\section{Datasets and Methodology}
\label{sec:analysis}

We now place observational constraints on the broken power-law SGWB spectrum we consider, with energy density spectrum given by Eq.~(\ref{eq:omegagwbrokenpowerlaw}), to examine whether it is possible to explain the NANOGrav signal while remaining consistent with a wide range of observational datasets, including those discussed in Sec.~\ref{sec:sources}. In our baseline analysis, we only consider CMB measurements alongside the constraints on the SGWB spectrum mentioned in Sec.~\ref{sec:sources} and the NANOGrav signal, which we treat as a genuine SGWB detection (with all the caveats discussed in Sec.~\ref{sec:intro}). In a later extended analysis, we further consider late-time cosmological measurements, to test whether they significantly improve cosmological parameter constraints, finding the answer to be negative.

In our most general analysis, we consider a ten-parameter model extending the standard six-parameter $\Lambda$CDM model by allowing four additional parameters to vary: the tensor-to-scalar ratio $r$, the pre-break tensor spectral index $n_T$ (not satisfying the inflationary consistency relation $r=-8n_T$, and hence independent of $r$), the break frequency $f_{\alpha}$, and the post-break tensor spectral index $\alpha$. Note that we consider purely adiabatic initial conditions, while restricting our analysis to a spatially flat Universe, and fixing the sum of the neutrino masses to $0.06\,{\rm eV}$, the minimum value allowed within the normal ordering, as the neutrino mass is known to correlate very weakly with inflationary parameters~\cite{Carbone:2010ik,Gerbino:2016sgw,Archidiacono:2016lnv}. Finally, we set the scalar and tensor pivot scales to $0.05\,{\rm Mpc}^{-1}$ and $0.01\,{\rm Mpc}^{-1}$ respectively.

Theoretical predictions for cosmological observables in the presence of the broken power-law SGWB spectrum are obtained through a modified version of the Boltzmann solver \texttt{CAMB}~\cite{Lewis:1999bs}. We add a module to \texttt{CAMB} which calculates the SGWB energy density as a function of wavenumber/frequency for a given choice of cosmological parameters, and the quantities $N_{\rm eff,BBN}^{\rm GW}$ and $N_{\rm eff,CMB}^{\rm GW}$ (which as argued earlier are approximately identical). Moreover, this module computes the amplitude of the SGWB strain power spectrum $A_{_{\rm CP}}$ at the NANOGrav reference frequency $f_{\rm yr}$, as parametrized by Eqs.~(\ref{PA1},\ref{PA2}), so that we can later treat it as a derived parameter.

We perform a Bayesian statistical analysis to constrain the model in question. We set flat priors on all cosmological parameters, except for $r$ and $f_{\alpha}$, with the prior ranges chosen to be wide enough as to not cut the associated marginalized posterior distributions where these are appreciably non-zero. For $r$, we instead set a prior flat in $\log_{10}r$. This choice makes the exploration of the parameter space much more efficient, particularly since explaining the NANOGrav signal while complying with upper limits from \textit{Planck} and BICEP2/Keck Array requires a small but non-zero $r$, as shown by one of us in Ref.~\cite{Vagnozzi:2020gtf}. These considerations suggest that a prior flat in $\log_{10}r$ may more correctly encapsulate our prior knowledge than a prior flat in $r$ does. Similarly, as $f_{\alpha}$ can span a wide range encompassing several orders of magnitude, we set a prior flat in $\log_{10}f_{\alpha}$, which better captures our prior information on the break frequency.~\footnote{More precisely, we actually set a prior flat in $\log_{10}k_{\alpha}$, with $k_{\alpha}$ the break wavenumber. However, given the proportionality relation between $k_{\alpha}$ and $f_{\alpha}$, a prior flat in $\log_{10}k_{\alpha}$ translates to a prior flat in $\log_{10}f_{\alpha}$.}

The posterior distributions for the cosmological parameters are sampled by means of Markov Chain Monte Carlo (MCMC) methods. We generate MCMC chains by using the cosmological MCMC sampler \texttt{CosmoMC}~\cite{Lewis:2002ah}, modified to interface itself with the modified version of \texttt{CAMB} discussed previously, and to include the additional observations (including NANOGrav) we further discuss below. The convergence of the generated MCMC chains is monitored via the Gelman-Rubin parameter $R-1$~\cite{Gelman:1992zz}, and we set the requirement $R-1<0.01$ in order for the chains to be considered converged.

As far as CMB data is concerned, we consider measurements of CMB temperature anisotropy and polarization power spectra, and their cross-spectra, from the \textit{Planck} 2018 legacy data release~\cite{Planck:2018vyg}. We combine the high-$\ell$ \texttt{Plik} likelihood for TT within the multipole range $30 \leq \ell \lesssim 2500$ with the TE and EE and low-$\ell$ TT-only likelihoods based on the \texttt{Commander} component-separation algorithm in pixel space within the multipole ranges $30 \leq \ell \lesssim 2000$ and $2 \leq \ell < 29$ respectively, and the low-$\ell$ EE-only \texttt{SimAll} likelihood in the multipole range $2 \leq \ell < 29$~\cite{Planck:2019nip}. We also make use of the likelihood for the \textit{Planck} CMB lensing power spectrum reconstructed from the temperature 4-point function~\cite{Planck:2018lbu}. Finally, we include the BICEP2/Keck Array B-mode polarization likelihood, which strongly constrains the amplitude of the tensor-to-scalar ratio at CMB scales~\cite{BICEP2:2015nss,BICEP2:2015xme,BICEP2:2018kqh} ($r_{0.05}<0.06$ at 95\%~C.L. when combined with \textit{Planck}). Note that the high-$\ell$ part of the CMB TT and EE likelihoods indirectly constrains $N_{\rm eff,CMB}^{\rm GW}$, through the effect of extra radiation on the damping tail and the eISW effect, as discussed in Sec.~\ref{subsec:cmb}.

Following earlier work~\cite{Giare:2020plo}, we treat the LIGO/Virgo upper limit on $\Omega_{\rm GW}(f_{\rm LV})$ reported in Eq.~(\ref{eq:ligolimit}) as a half-Gaussian prior on the amplitude of the SGWB energy density at the frequency in question, which is thus treated as a derived parameter. We choose to use this constraint and not the most recent ones~\cite{KAGRA:2021kbb} to obtain the most conservative results possible.

We instead treat the BBN limit as a hard upper limit on $N_{\rm eff,BBN}^{\rm GW}<0.4$. However, we stress that this information is essentially redundant, given that within the assumed SGWB spectrum $N_{\rm eff,CMB}^{\rm GW} \approx N_{\rm eff,BBN}^{\rm GW}$, and the adopted CMB likelihoods already set a tighter limit on $N_{\rm eff,CMB}^{\rm GW}$, through the effect of the latter on the damping tail and the eISW effect.

For what concerns the NANOGrav signal, we approximate this as a multivariate Gaussian on the two derived parameters $\log_{10}A_{_{\rm CP}}$ and $\gamma_{_{\rm CP}}$, defined in Eqs.~(\ref{PA1},\ref{PA2}), and evaluated at the reference frequency $f_{\rm yr}$ relevant for NANOGrav. Specifically, if we denote by $\boldsymbol{x}(\boldsymbol{\theta})$ the vector of derived parameters $\boldsymbol{x} \equiv \{\log_{10}A_{_{\rm CP}}(\boldsymbol{\theta}),\gamma_{_{\rm CP}}(\boldsymbol{\theta})\}$, where $\boldsymbol{\theta}$ is the 10-dimensional vector of cosmological parameters, we approximate the NANOGrav log-likelihood as being given by:
\begin{eqnarray}
\ln{\cal L}_{\rm NANOGrav}(\boldsymbol{\theta}) = -\frac{\chi^2_{\rm NANOGrav}(\boldsymbol{\theta})}{2}\,,
\label{eq:loglikelihoodnanograv}
\end{eqnarray}
where $\chi^2_{\rm NANOGrav}$ is given by:
\begin{eqnarray}
\chi^2_{\rm NANOGrav} = (\boldsymbol{x}(\boldsymbol{\theta})-\boldsymbol{\mu})^T\boldsymbol{\Sigma}^{-1}(\boldsymbol{x}(\boldsymbol{\theta})-\boldsymbol{\mu})\,.
\label{eq:chi2}
\end{eqnarray}
In Eq.~(\ref{eq:chi2}), $\boldsymbol{\mu} \approx (-15.2,5.3)$ is the vector of mean values for the derived parameters $\log_{10}A_{_{\rm CP}}$ and $\gamma_{_{\rm CP}}$, and $\boldsymbol{\Sigma}$ is the covariance matrix (with $^T$ denoting the transpose operation), which we numerically determine to be $\boldsymbol{\Sigma} \approx \left ( \begin{smallmatrix}0.1 & -0.182\\ -0.182 & 0.4\end{smallmatrix} \right ) $. To test the goodness of the multivariate Gaussian likelihood approximation in Eqs.~(\ref{eq:loglikelihoodnanograv},\ref{eq:chi2}), with the approximate values of $\boldsymbol{\mu}$ and $\boldsymbol{\Sigma}$ we determined, we draw $10000$ random samples from the probability distribution function defined by ${\cal L}(\boldsymbol{x})$, and verify that the associated 68\% and 95\%~C.L. constraints on $\log_{10}A_{_{\rm CP}}$ and $\gamma_{_{\rm CP}}$ fall to very good approximation within the corresponding $1\sigma$ and $2\sigma$ contours given in the right panel of Fig.~1 of Ref.~\cite{NANOGrav:2020bcs}. While this is of course an approximation, for the purposes of our analysis it provides a sufficiently precise compression of the information content of the tentative NANOGrav detection.~\footnote{Alternatively, we note that the $\log_{10}A_{_{\rm CP}}$ and $\gamma_{_{\rm CP}}$ posterior samples for the NANOGrav signal are available at: \url{https://github.com/nanograv/12p5yr_stochastic_analysis}}. Finally, while in Eqs.~(\ref{eq:loglikelihoodnanograv},\ref{eq:chi2}) we have retained the dependence on the full set of cosmological parameters $\boldsymbol{\theta}$, we stress that to very good approximation the NANOGrav likelihood mostly depends on the tensor-to-scalar ratio $r$ (with a minor dependence on $A_s$, as well as on $\Omega_m$ and $H_0$, themselves depending on the fundamental parameters $\omega_b$, $\omega_c$, and $\theta_s$, all of which are very well constrained by the CMB), and the tensor spectral index $n_T$, with $\gamma_{_{\rm CP}}$ depending only on $n_T$ since $\gamma_{_{\rm CP}}=5-n_T$ as discussed earlier. In any case, we numerically determine $\log_{10}A_{_{\rm CP}}$ and $\gamma_{_{\rm CP}}$ at each step of our MCMC analysis.

Our baseline analysis combines CMB data from \textit{Planck} and BICEP2/Keck Array, constraints on the SGWB energy density from LIGO/Virgo and BBN, and the tentative NANOGrav detection. We refer to this combination of likelihoods as ``\textit{\textbf{base}}''.

At a later stage, we combine the \textit{base} dataset with additional late-time distance and expansion rate measurements. In particular, we consider Baryon Acoustic Oscillation (BAO) measurements from the 6dFGS~\cite{Beutler:2011hx}, SDSS DR7 MGS~\cite{Ross:2014qpa}, and BOSS DR12 samples~\cite{BOSS:2016wmc},~\footnote{When this project was initiated, the likelihood for the completed lineage of SDSS experiments, including in particular the legacy eBOSS measurements~\cite{eBOSS:2020yzd}, was not yet publicly available. In any case, we expect that the inclusion of the latter measurements would not quantitatively alter our results, given our a posteriori finding that the inclusion of late-time measurements returns cosmological constraints which are virtually identical to those obtained from the \textit{base} dataset combination alone.} distance moduli measurements from the \textit{Pantheon} Type Ia Supernovae (SNeIa) compilation~\cite{Scolnic:2017caz}, and measurements of the expansion rate $H(z)$ from cosmic chronometers (CC), using the relative ages of massive, early-time, passively evolving galaxies as first proposed in Ref.~\cite{Jimenez:2001gg}, and reported in Refs.~\cite{Jimenez:2003iv,Stern:2009ep,Moresco:2012by,Zhang:2012mp,Moresco:2015cya,Moresco:2016mzx,Ratsimbazafy:2017vga,Borghi:2021rft} (see Tab.~1 in Ref.~\cite{Vagnozzi:2020dfn}).~\footnote{We have conservatively chosen to omit the measurements reported in Ref.~\cite{Simon:2004tf}, given the concerns on these measurements recently raised in Ref.~\cite{Kjerrgren:2021zuo}.} We refer to the combination of BAO, SNeIa, and CC measurements as ``\textit{\textbf{late}}''.

Our goal in combining the \textit{base} and \textit{late} dataset combinations is to check whether the latter can improve the determination of the SGWB parameters, in particular $\alpha$ and $f_{\alpha}$. Since BAO, SNeIa, and CC measurements do not directly constrain the SGWB spectrum, one might legitimately wonder why to expect an improvement in first place. The reason is that these measurements help improving constraints on parameters which may be degenerate with the SGWB ones, and in particular the tensor-to-scalar ratio $r$. Moreover, as we stated earlier, $\log_{10}A_{_{\rm CP}}$ carries some amount of dependence (albeit very weak) on $\Omega_m$ and $H_0$. In any case, we find a posteriori that adding the \textit{late} dataset combination to the \textit{base} one does not improve the constraints on the SGWB parameters, implying that the aforementioned degeneracies play a very marginal role in our analysis.

\begin{table}[t]
\setlength{\tabcolsep}{0.5em}
\renewcommand{\arraystretch}{1.5}
\centering
\begin{tabular}{cccc}
\hline 
\hline
{$f_{\alpha}$, $r$} & 
{$f_{\alpha}$ [Hz]} &
{$r$} & 
{$n_T$} \\
\hline
Fixed   & \vline \; $10^{-6} / 10^{-3}$ \; \vline &  $10^{-6}$  & \vline \; $1.31 \pm 0.03$  \\
\hline
Fixed   & \vline \; $10^{-6} / 10^{-3}$ \; \vline &  $10^{-3}$  & \vline \; $0.98 \pm 0.03$  \\
\hline
Fixed   & \vline \; $10^{-6} / 10^{-3}$ \; \vline &  $0.06$  & \vline \; $0.79 \pm 0.03$  \\
\hline
Free   & \vline  \;  $~< 10^{-0.7}~$ \; \vline & $  >10^{-6.6}$  & \vline \; $0.97 \pm 0.19$  \\
\hline 
\hline
\end{tabular}
\caption{
Constraints on the SGWB parameters using the \textit{base} dataset combination. In the first three analyses $f_{\alpha}$ and $r$ are fixed, while in the fourth analysis these two parameters are varied (while fixing $\alpha=-3$). The constraints obtained within the latter analysis are very stable against the further introduction of $\alpha$ as a free parameter. Quoted intervals correspond to 68\%~C.L. intervals, whereas quoted upper/lower limits correspond to 95\%~C.L. upper/lower limits.}
\label{tab:results}
\end{table}

In our analysis, we choose to proceed in two distinct steps, to better understand to what extent the data is actually sensitive to the beyond-$\Lambda$CDM parameters (in particular $f_{\alpha}$, $r$, and $\alpha$), and to aid the convergence of our MCMC chains. In a first instance, we fix $f_{\alpha}$ and $r$ to certain sets of values, and examine the resulting constraints on $n_T$ and $\alpha$ (as well as the other cosmological parameters). We find that the data is not strongly sensitive to $\alpha$, so in the second instance we fix this parameter, and vary all the other 9 cosmological parameters. However, we perform a final sanity check to ensure that varying $\alpha$ does not significantly affect our results, finding this to indeed be the case. We would also like to stress that, as we are taking into account rather negative values of $\alpha$, considering the more stringent constraints on the SGWB amplitude given by the more recent LIGO/VIRGO results~\cite{KAGRA:2021kbb} would not qualitatively affect our results.

\section{Results and discussion}
\label{sec:results}

\begin{figure*} [t!]
\centering
\includegraphics[width=0.8\linewidth]{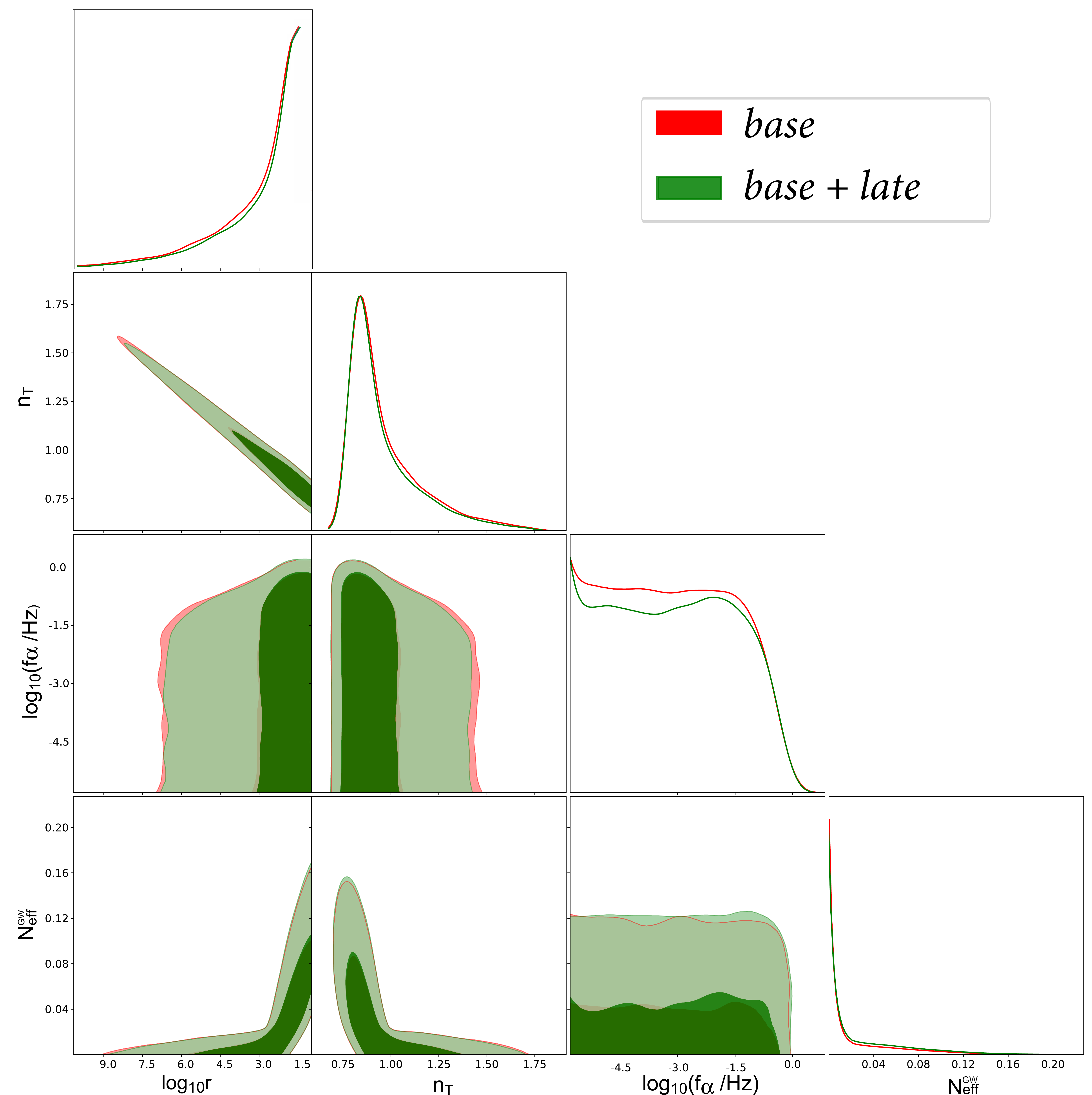}
\caption{Triangular plot showing 2D joint and 1D marginalized posterior probability distributions for $\log_{10}r$, $n_T$, $\log_{10}(f_{\alpha}/{\rm Hz})$, and $N_{\rm eff}^{\rm GW}$ from the \textit{base} (red contours) and \textit{base}+\textit{late} (green contours) dataset combinations, with the former only including CMB and GW observations (including the tentative NANOGrav signal), and the latter also including late-time distance and expansion rate BAO, SNeIa, and CC measurements. The corresponding constraints have been obtained by varying all the cosmological parameters except for the post-break frequency $\alpha$, fixed to $\alpha=-3$. It is clear that the inclusion of late-time measurements has a negligible effect on the resulting constraints.}
\label{fig:tri_plot}
\end{figure*}

As stated above, we start by understanding to what extent the data is sensitive to the break frequency $f_{\alpha}$ and the tensor-to-scalar ratio $r$. We do so by fixing these parameters, and considering the \textit{base} dataset combination. Specifically, we fix $f_{\alpha}$ to the values $10$, $10^{-3}$, and $10^{-6}\,{\rm Hz}$, and $r$ to the values $0.06$ (saturating constraints from \textit{Planck} and BICEP2/Keck Array within a $\Lambda$CDM+$r$ model), $10^{-3}$, and $10^{-6}$. In particular, the break frequency is chosen to lie between the NANOGrav and LIGO/Virgo frequencies. The results of these tests are reported in Tab.~\ref{tab:results}.

We find, as expected, that the smaller the (fixed) value of $r$, the larger the inferred value of $n_T$. This is not unexpected, and in line with the findings in Ref.~\cite{Vagnozzi:2020gtf}: if the SGWB amplitude on CMB frequencies is smaller, the tensor tilt will necessarily have to be larger in order for the SGWB amplitude to grow more quickly with increasing frequency and match the amplitude of the NANOGrav signal at the respective frequency. This behavior is clearly shown in Tab.~\ref{tab:results}, where the first three rows report the inferred value of $n_T$ as a function of the (fixed) value of $r$, and we can see that the former decreases (increases) while the latter increases (decreases).

For what concerns the break frequency $f_{\alpha}$ we find that, as long as this parameter lies below a certain value, it has little effect on our results: in other words, within this range, the likelihood is roughly flat along the $f_{\alpha}$ direction. Larger frequencies, however, are unable to match the amplitude of the NANOGrav signal while respecting $N_{\rm eff}^{\rm GW}$ constraints (again in line with what was found earlier in Ref.~\cite{Vagnozzi:2020gtf}, focusing on the reheating temperature $T_{\rm rh}$ which can be related to $f_{\alpha}$). For instance we find that $f_{\alpha}=10\,{\rm Hz}$ requires a very low value of $A_{_{\rm CP}} \lesssim 10^{-16}$ in order not to run afoul of $N_{\rm eff}^{\rm GW}$ constraints, and is thus unable to explain the NANOGrav signal, therefore being disfavored by our analysis.

Here, we have allowed $\alpha$ to vary, requiring this parameter to be negative so that the high-frequency part of the broken SGWB spectrum is red. We find, as expected, that we can only set an upper limit on this parameter. This is not surprising since, leaving aside $N_{\rm eff}^{\rm GW}$ constraints for the moment, an arbitrarily large negative $\alpha$ can be invoked in order for the high-frequency part of the SGWB spectrum to be consistent with LIGO/Virgo's upper limit. For example, for $r=10^{-6}$ we find $\alpha<-2.8$ at 95\%~C.L., whereas this limit changes to $\alpha<-0.9$ and $\alpha<-0.7$ for $r=10^{-3}$ and $r=0.06$ respectively. It is worth mentioning that these results, alongside those reported in Tab.~\ref{tab:results}, can easily be seen to satisfy the relation $\alpha=-2+n_T$, expected from models with non-instantaneous reheating: this confirms that in light of our observational constraints, the broken power-law SGWB spectrum adopted can be interpreted in the context of these models~\cite{Kuroyanagi:2014nba,Cook:2015vqa,Kuroyanagi:2020sfw}.

In the second part of our analysis, we vary all cosmological parameters (including $f_{\alpha}$ and $r$), except for $\alpha$. The reason is that we expect the likelihood to be roughly flat along the $\alpha$ direction: in fact, any sufficiently negative $\alpha$ will fit the data, and more precisely the upper limit set by LIGO/Virgo, equally well. If we do not commit to any specific theory, there is in principle no well-motivated theoretical lower limit to the value $\alpha$ can take within our phenomenological parametrization of Eq.~(\ref{eq:omegagwbrokenpowerlaw}). Since we are mostly interested in constraints on the pre-break part of the SGWB spectrum, \textit{i.e.} the one responsible for explaining the NANOGrav signal, we decide to fix $\alpha=-3$. This arbitrary value is sufficiently negative as to not impact our results: in other words, as per the results of the first part of our analysis, this choice guarantees that the broken power-law SGWB spectrum will be consistent with the LIGO/Virgo upper limit for any choice of parameters explaining NANOGrav.

We present a triangular plot of the constraints on the main parameters of interest in Fig.~\ref{fig:tri_plot}, including the derived parameter $N_{\rm eff}^{\rm GW}$, which quantifies the contribution of the SGWB spectrum to the radiation energy density. We find that this parameter is very strongly constrained to $N_{\rm eff}^{\rm GW} \lesssim 0.11\,[0.28]$ at 68 [95] \%~C.L., given the impact of extra radiation on the damping tail and eISW effect, both tightly constrained by CMB data. Let us stress that this result strictly depends on the assumptions made, including the standard $\eta$-dependence in the transfer function. This implies that our results may not in general be applicable to models which, while being compatible with the broken power-law SGWB spectrum, have a different $\eta$-dependence in the transfer function (as these may provide different $N_{\rm eff}^{\rm GW}$ constraints). However, to understand whether this is the case requires committing to a specific non-standard model of inflation, whereas here we have taken a model-agnostic phenomenological stand.

Moreover, we obtain a detection of non-zero $r$, with the lower limit $r>2.5 \times 10^{-7}$ at 95\%~C.L., consistent with earlier findings in Ref.~\cite{Vagnozzi:2020gtf}, as smaller values of $\alpha$ would require too large a tilt, disfavored by the NANOGrav signal (and in particular the constraints on $\gamma_{_{\rm CP}}$). Our analysis also confirms that we require an extremely blue spectrum for frequencies below the break, as we infer $n_T=0.97 \pm 0.19$.~\footnote{This is broadly consistent with the earlier findings of Ref.~\cite{Vagnozzi:2020gtf}, which however did not perform a full Bayesian analysis but just a parameter scan.} In addition, we find an upper limit on the break frequency, with $f_{\alpha}<0.2\,{\rm Hz}$ at 95\%~C.L.: this is again as expected given the results of our previous simplified analysis with $f_{\alpha}$ and $r$ fixed, as larger values of $f_{\alpha}$ would run afoul of constraints on $N_{\rm eff}^{\rm GW}$. 

The results discussed so far were obtained using the \textit{base} dataset combination. We include the \textit{late} dataset combination to see whether this improves constraints on any of the cosmological parameters. We find that this mostly leads to improvements in the inferred values of non-SGWB derived parameters, such as $\Omega_m$ and $H_0$. For instance, the inferred 68\%~C.L. constraints on $H_0$ shift from $H_0=(67.38 \pm 0.53)\,{\rm km}\,{\rm s}^{-1}\,{\rm Mpc}^{-1}$ to $H_0=(67.78 \pm 0.40)\,{\rm km}\,{\rm s}^{-1}\,{\rm Mpc}^{-1}$. Therefore, unlike what was found by some of us using earlier data in Ref.~\cite{Graef:2018fzu}, a blue GW spectrum is unable to solve or even alleviate the Hubble tension \cite{Graef:2017cfy}. It is known that a larger value of $N_{\rm eff}$ is in principle a viable way of alleviating the $H_0$ tension, as a larger radiation component reduces the sound horizon, allowing for a larger value of $H_0$ (see e.g. Refs.~\cite{DiValentino:2016hlg,Bernal:2016gxb,Benetti:2017gvm,Benetti:2017juy,Mortsell:2018mfj,DEramo:2018vss,Guo:2018ans,Kreisch:2019yzn,Vagnozzi:2019ezj,Ballesteros:2020sik,DiValentino:2020zio,RoyChoudhury:2020dmd,Brinckmann:2020bcn,Seto:2021xua,DiValentino:2021izs}). However, the updated datasets we have used here severely constrain $N_{\rm eff}^{\rm GW}$ (primarily through its impact on Silk damping~\cite{Silk:1967kq} and the eISW effect~\cite{Hou:2011ec,Cabass:2015xfa,Kable:2020hcw,Vagnozzi:2021gjh}), therefore strongly limiting the possibility of appreciably raising $H_0$ within this model. We also find that constraints on the SGWB parameters are virtually unaffected, only improving very slightly (and to an extent which is in principle compatible with a statistical fluctuation given the convergence level of the MCMC chains), implying that degeneracies with non-SGWB parameters whose determination is improved by including late-time datasets plays no major role in our analysis.

Finally, as a sanity check, we perform a run where we vary all 10 cosmological parameters including $\alpha$, for which we set a flat prior with a purely indicative lower prior limit of $-10$. We find that the constraints on the other 9 cosmological parameters are very stable against the introduction of $\alpha$ as a free parameters. Moreover, we infer a rather loose upper limit on $\alpha<-0.6$ at 95\%~C.L., limit which was well satisfied in our previous analysis where we chose to fix $\alpha=-3$. This final analysis confirms that the likelihood is roughly flat in the $\alpha$ direction provided $\alpha$ is sufficiently negative, and that fixing this parameter to a sufficiently negative value does not have an appreciable effect on our results, which are mainly constraining the parameters governing the pre-break part of the broken power-law SGWB spectrum (parameters which we are also most interested in).

Overall, our results show that the broken power-law SGWB spectrum can be brought in agreement with a wide multi-frequency range of observations and constraints, ranging from CMB and BBN to interferometers, while at the same time potentially explaining the tentative NANOGrav detection. In particular, the break in the power spectrum, which can be justified by several well-motivated fundamental physics scenarios as discussed in Sec.~\ref{subsec:brokenpowerlaw}, can significantly reduce the SGWB contribution to the radiation energy density in the early Universe as captured by $N_{\rm eff}^{\rm GW}$, which would otherwise be very large within the simple power-law model~\cite{Graef:2018fzu,Vagnozzi:2020gtf}, while allowing the SGWB spectrum to remain in agreement with constraints from LIGO/Virgo. All of this is possible for a wide range of break frequencies $f_{\alpha}$ and post-break tilt $\alpha$, as long as the latter is negative so that the post-break SGWB spectrum is red, while being blue for lower frequencies (to explain the NANOGrav signal while remaining in agreement with upper limits on the tensor-to-scalar ratio on CMB scales).

\section{Conclusions and Prospects}
\label{sec:conclusions}

 \begin{figure*}[t!]
\centering
\includegraphics[width=0.8\linewidth]{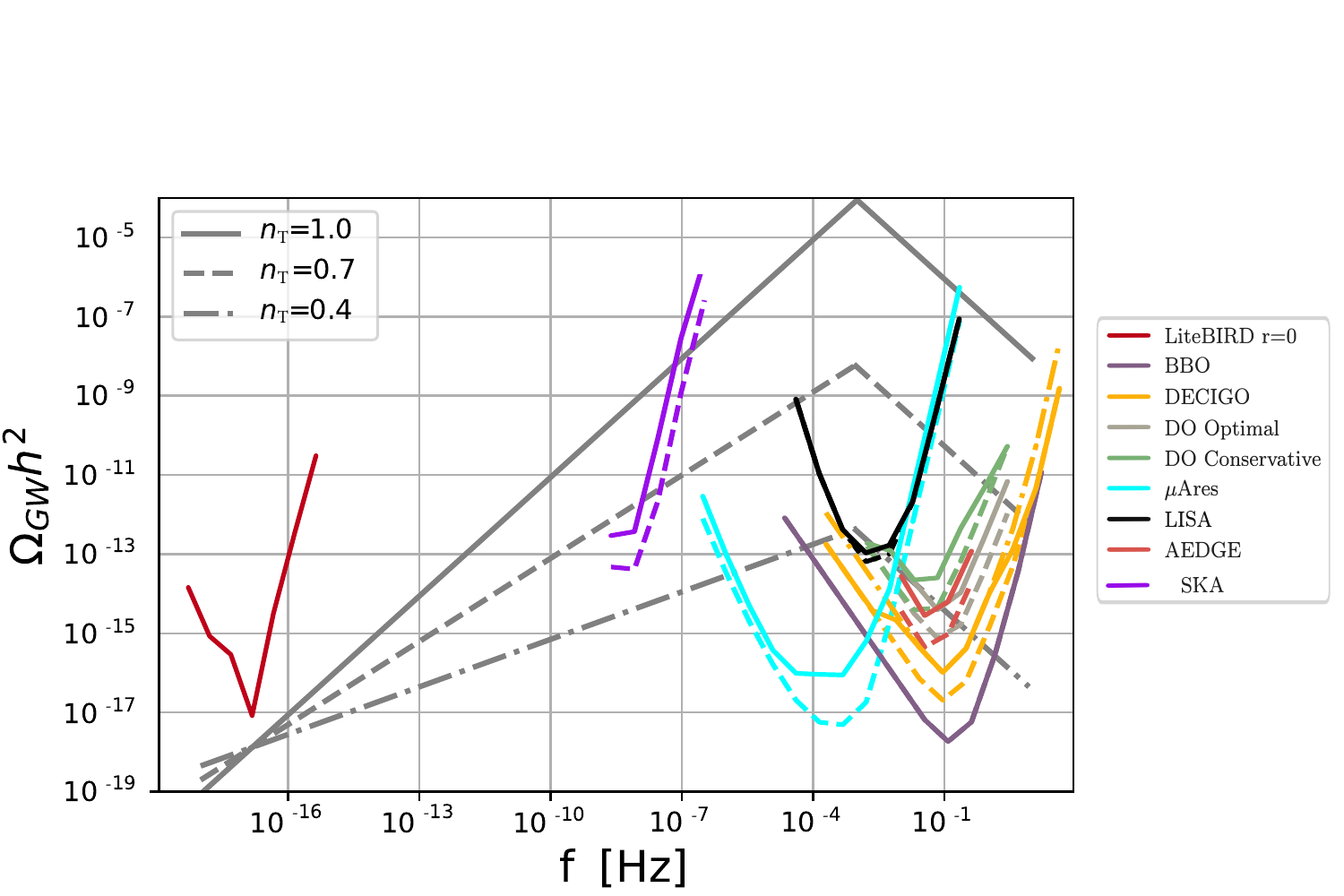}
\caption{Benchmark examples of the broken power-law SGWB spectrum considered in this work, with $r=10^{-4}$, $f_{\alpha}=10^{-3}\,{\rm Hz}$, $\alpha=-1$, and three different values of $n_T=1.0$ (grey solid line, consistent with the NANOGrav signal), $0.7$ (grey dashed line, marginally consistent with the NANOGrav signal), and $0.4$ (grey dashed-dotted line, inconsistent with the NANOGrav signal). The colored curves indicate the expected sensitivity curves of future experiments (as per the color coding) briefly discussed later in Sec.~\ref{sec:conclusions}, with (solid curves) and without (dashed curves) the contribution of astrophysical foregrounds. Figure adapted from Fig.~8 of Campeti, Komatsu, Poletti \& Baccigalupi, ``\textit{Measuring the spectrum of primordial gravitational waves with CMB, PTA and laser interferometers}'', JCAP 01 (2021) 012, published 8 January 2021~\cite{Campeti:2020xwn}. \textcopyright IOP Publishing Ltd and Sissa Medialab.  Reproduced by permission of IOP Publishing. All rights reserved.}
\label{fig:omegaGW-vs-f}
\end{figure*}

NANOGrav's possible SGWB detection~\cite{NANOGrav:2020bcs} could constitute a significant milestone towards our understanding of the physics of the very early Universe. While the most natural candidate for a signal in the nHz range is an astrophysical SGWB arising from the merger of SMBHs, it is worth investigating whether NANOGrav's signal could instead be cosmological in nature. One of the best motivated candidates in this sense is a SGWB generated during an early period of inflation. In this paper, we have re-examined the possibility that NANOGrav may have detected primordial GWs, arising from inflation or early-Universe alternatives to inflation. In particular, we have extended earlier works by going beyond the overly simplistic assumption of a pure power-law SGWB spectrum, considering constraints on the SGWB amplitude from a variety of sources, and performing a more complete analysis adopting various precision cosmological datasets.

We have considered a broken power-law SGWB spectrum [Eq.~(\ref{eq:omegagwbrokenpowerlaw})], with the spectral index changing from $n_T$ to $\alpha$ above a characteristic break frequency $f_{\alpha}$. While we have taken a phenomenological stance, without committing to any specific early-Universe model, we have argued (see Sec.~\ref{subsec:brokenpowerlaw}) that such a phenomenological parametrization is actually flexible enough to cover a wide range of interesting early Universe models. These include scenarios altering the SGWB transfer function such as inflationary models with low-scale reheating and/or a non-standard background evolution following reheating, and late-time entropy injection, as well as scenarios altering the primordial SGWB power spectrum, such as models involving particle production, or alternatives to inflation.

This particular choice of SGWB spectrum allows us to explain the tentative NANOGrav detection in the ${\cal O}({\rm nHz})$ range while complying with upper limits on the tensor-to-scalar ratio on CMB scales [$f \ll {\cal O}(10^{-15})\,{\rm Hz}$], and on the SGWB amplitude on interferometer scales [$f \sim {\cal O}(10)\,{\rm Hz}$]. In particular, we require a spectrum which is blue for frequencies below the break, and red above, with the break frequency lying between the PTA and interferometer ranges. The break helps \textit{a)} ensuring consistency with LIGO/Virgo's upper limits (a requirement which had been missed earlier in Ref.~\cite{Vagnozzi:2020gtf}), and \textit{b)} suppressing the SGWB contribution to the radiation energy density in the early Universe (as captured by the parameter $N_{\rm eff}^{\rm GW}$), which in turn is key in order not to run afoul of CMB and BBN constraints.

We constrain the broken power-law SGWB spectrum using a wide range of cosmological and astrophysical datasets, including CMB (\textit{Planck} and BICEP2/Keck Array) and BBN, limits from LIGO/Virgo, the tentative NANOGrav signal which we interpret as a genuine detection, and late-time measurements of the expansion history. This dataset combination strongly constrains the GW contribution to the radiation energy density, with the 95\%~C.L. upper limit $N_{\rm eff}^{\rm GW} \lesssim 0.11$. At the same time, we find that explaining the NANOGrav detection requires a very blue spectrum below the break, with $n_T=0.97 \pm 0.19$, and of course a non-zero value for the tensor-to-scalar ratio, with the 95\%~C.L. lower limit $r>2.5 \times 10^{-7}$, perfectly in agreement with upper limits from the non-observation of a primordial B-mode polarization signal. We find that the model is able to satisfy BBN and LIGO/Virgo limits for a wide range of values for the post-break tilt $\alpha<0$, while we find the 95\%~C.L. upper limit on the break frequency $f_{\alpha}<0.2\,{\rm Hz}$.

The constraints we obtained on the SGWB spectrum parameters are still rather loose, primarily due to the sparsity and still limited precision of the available datasets constraining the SGWB across 20 decades in frequency (although we recall that CMB and BBN carry integrated sensitivity to the SGWB spectrum throughout a wide frequency range). Several next-generation experiments will fill the gap between PTA and interferometer frequencies~\cite{Campeti:2020xwn} as shown in Fig.~\ref{fig:omegaGW-vs-f}. Here we plot three benchmark example SGWB spectra such as the one we considered (with three values of $n_T$ respectively consistent, marginally consistent, and inconsistent with the NANOGrav signal), alongside the expected noise curves of some of these upcoming experiments (with and without the contribution of astrophysical foregrounds, whose inclusion lowers the expected sensitivity level of these experiments), as obtained in the very detailed analysis of Ref.~\cite{Campeti:2020xwn}. It is clear that the sensitivities of these experiments (featuring improvements of up to 2-3 orders of magnitude compared to existing ones) and their expanded band width will significantly aid our ability to constrain the SGWB spectrum considered, particularly for what concerns the break frequency and tensor tilt below and above the break.

Examples of upcoming surveys which will be extremely helpful towards further testing the phenomenological model we considered include but are not limited to the following (in brackets the frequency range these will probe): the SKA, which is expected to detect thousands of millisecond pulsars ($10^{-9}$-$10^{-7}\,{\rm Hz}$)~\cite{Weltman:2018zrl}, $\mu$Ares ($10^{-6}$-$10^{-2}$Hz)~\cite{Sesana:2019vho}, LISA ($10^{-4}$-$10^{-1}\,{\rm Hz}$)~\cite{LISA:2017pwj}, BBO~\cite{Crowder:2005nr} and DECIGO ($10^{-4}$-$10\,{\rm Hz}$)~\cite{Seto:2001qf}, DO ($10^{-3}$-$10\,{\rm Hz}$)~\cite{Sedda:2019uro}, AEDGE ($10^{-2}$-$10^0\,{\rm Hz}$)~\cite{AEDGE:2019nxb}, as well as the Einstein Telescope ($10^0$-$10^3\,{\rm Hz}$)~\cite{Maggiore:2019uih}. Binary resonance, for instance through laser ranging of the Moon, artificial satellites around the Earth, and timing of binary pulsars, will be able to probe a wide frequency range encompassing the ${\cal O}(\mu{\rm Hz})$ gap between the most sensitive bands of PTA and interferometers~\cite{Blas:2021mpc,Blas:2021mqw}. Future CMB missions such as the space-based LiteBIRD~\cite{Matsumura:2013aja} and the ground-based CMB-S4~\cite{CMB-S4:2016ple} and Simons Observatory~\cite{SimonsObservatory:2018koc,SimonsObservatory:2019qwx} will be able to probe values of the tensor-to-scalar ratio as low as ${\cal O}(10^{-3})$ through the associated primordial B-mode polarization signal. The combination of all these probes will enable a precise characterization of the SGWB spectrum across 21 decades in frequency,~\footnote{Including BBN information will potentially extend the range to 29 decades, given our assumption of $f_{\rm UV} \sim 10^8\,{\rm Hz}$, although we stress once more that our results are insensitive to this choice.}, potentially allowing to distinguish between different theoretical origins for the phenomenological spectrum we have assumed, and in particular whether the latter is sourced by quantum vacuum fluctuations in the metric tensor, sources such as production of gauge fields coupled to the inflaton, or primordial scenarios alternative to inflation.

While a detection of spatial quadrupolar correlations is required for the NANOGrav signal to be confirmed as a genuine SGWB detection, we believe there is nonetheless reason to be cautiously optimistic, particularly given PPTA and EPTA's tentative detections in a similar frequency range~\cite{Goncharov:2021oub,Chen:2021rqp}. With these caveats in mind, our work reinforces the possibility (also discussed in previous works~\cite{Vagnozzi:2020gtf,Kuroyanagi:2020sfw,Li:2021htg}) that the NANOGrav signal may be primordial in origin, associated to early-Universe scenarios such as inflation or alternatives thereto. Fitting a wide range of precision cosmological and astrophysical datasets within a phenomenological (but representative of several well-motivated theoretical scenarios) broken power-law SGWB spectrum~\cite{Kuroyanagi:2014nba}, we have confirmed that this interpretation requires several deviations from the standard picture (e.g.\ instantaneous reheating and standard post-reheating background evolution), and reinforced the importance of multi-frequency GW observations towards better understanding the dynamics of the primordial Universe. While we await a definitive confirmation of NANOGrav's signal, there are several interesting avenues for follow-up work, particularly related to connections to more specific primordial scenarios (inflationary or not) and observational implications thereof, such as primordial BHs, observable non-Gaussianity, or connections to the possible non-tensorial nature of the NANOGrav signal~\cite{Chen:2021wdo,Chen:2021ncc,Wu:2021kmd,NANOGrav:2021ini}. We defer the investigation of these and related issues to future work.

\section*{Acknowledgments}
MB is supported by the Istituto Nazionale di Fisica Nucleare (INFN), Sezione di Napoli, iniziativa specifica QGSKY. L.L.G  is supported by the Conselho Nacional de  Desenvolvimento Cientıfico e Tecnologico (CNPq), Grant No. 307052/2019-2, and by the Fundaç\~{a}o Carlos Chagas Filho de Amparo \'{a} Pesquisa do Estado do Rio de Janeiro--FAPERJ, Grant No. E-26/201.297/2021. S.V. is supported by the Isaac Newton Trust and the Kavli Foundation through a Newton-Kavli Fellowship, and by a grant from the Foundation Blanceflor Boncompagni Ludovisi, n\'{e}e Bildt. S.V. acknowledges a College Research Associateship at Homerton College, University of Cambridge. We acknowledge the use of the \texttt{CAMB}~\cite{Lewis:1999bs} and \texttt{CosmoMC}~\cite{Lewis:2002ah} codes. We acknowledge the use of computational facilities provided by the High Performance Data Center (DCON) of the Observat\'{o}rio Nacional.

\bibliographystyle{apsrev4-1}
\bibliography{NANOGrav.bib}

\end{document}